\documentclass[aps,prl,twocolumn,epsfig,superscriptaddress,showpacs,amsmath,amsfonts,amssymb,floatfix,longbibliography]{revtex4-1}
\usepackage{graphicx}
\usepackage{bm}
\usepackage{epstopdf}
\usepackage{color}
\usepackage{amsmath}
\usepackage[english]{babel}
\usepackage{float}
\usepackage{multibib}

\newcommand{\eff}{\textnormal{\tiny \textsc{E}}}
\newcommand{\act}{\textnormal{\tiny \textsc{A}}}
\newcommand{\pas}{\textnormal{\tiny \textsc{P}}}
\newcommand{\drg}{\textnormal{\tiny \textsc{D}}}
\newcommand{\scr}[1]{\textnormal{\tiny \textsc{#1}}}

\begin{document}

\title{Theory of Coulomb Drag in Spatially Inhomogeneous Materials}

\author{Derek Y.H. Ho}
\affiliation{Centre for Advanced 2D Materials and Department of Physics, National University of Singapore, 2 Science Drive 3, 117551, Singapore}
\author{Indra Yudhistira}
\affiliation{Centre for Advanced 2D Materials and Department of Physics, National University of Singapore, 2 Science Drive 3, 117551, Singapore}
\author{Ben Yu-Kuang Hu}
\affiliation{Department of Physics, University of Akron, Akron, Ohio 44325-4001, USA}
\author{Shaffique Adam}
\affiliation{Centre for Advanced 2D Materials and Department of Physics, National University of Singapore, 2 Science Drive 3, 117551, Singapore}
\affiliation{Yale-NUS College, 6 College Avenue East, 138614, Singapore}

\begin{abstract}

Coulomb drag between parallel two-dimensional electronic layers is an excellent tool for the study of electron-electron interactions. In actual experiments, the layers display spatial charge density fluctuations due to imperfections such as external charged impurities. However, at present a systematic way of taking these inhomogeneities into account in drag calculations has been lacking, making the interpretation of experimental data problematic. On the other hand, there exists a highly successful and widely accepted formalism describing transport within single inhomogeneous layers known as effective medium theory. 
In this work, we generalize the standard effective medium theory to the case of Coulomb drag between two inhomogeneous sheets and demonstrate that inhomogeneity in the layers has a strong impact on drag transport. In the case of exciton condensation between the layers, we show that drag resistivity takes on a value determined by the amplitude of density fluctuation. Next we consider drag between graphene sheets, in which the existence of spatial charge density fluctuations is well-known. We show that these inhomogeneities play a crucial role in explaining existing experimental data. In particular, the temperature dependence of the experimentally observed peaks in drag resistivity can only be explained by taking the layer density fluctuations into account. 
We also propose a method of extracting information on the correlations between the inhomogeneities of the layers. 
The effective medium theory of Coulomb drag derived here is general and applies to all two-dimensional materials.

\end{abstract}

\date{\today}
\maketitle

\section{I. INTRODUCTION}

\begin{figure}[h!]
	\begin{center}$
		\begin{array}{c}
		\includegraphics[trim=0cm 3cm 0cm 2cm,clip=true,height=! ,width=9cm] {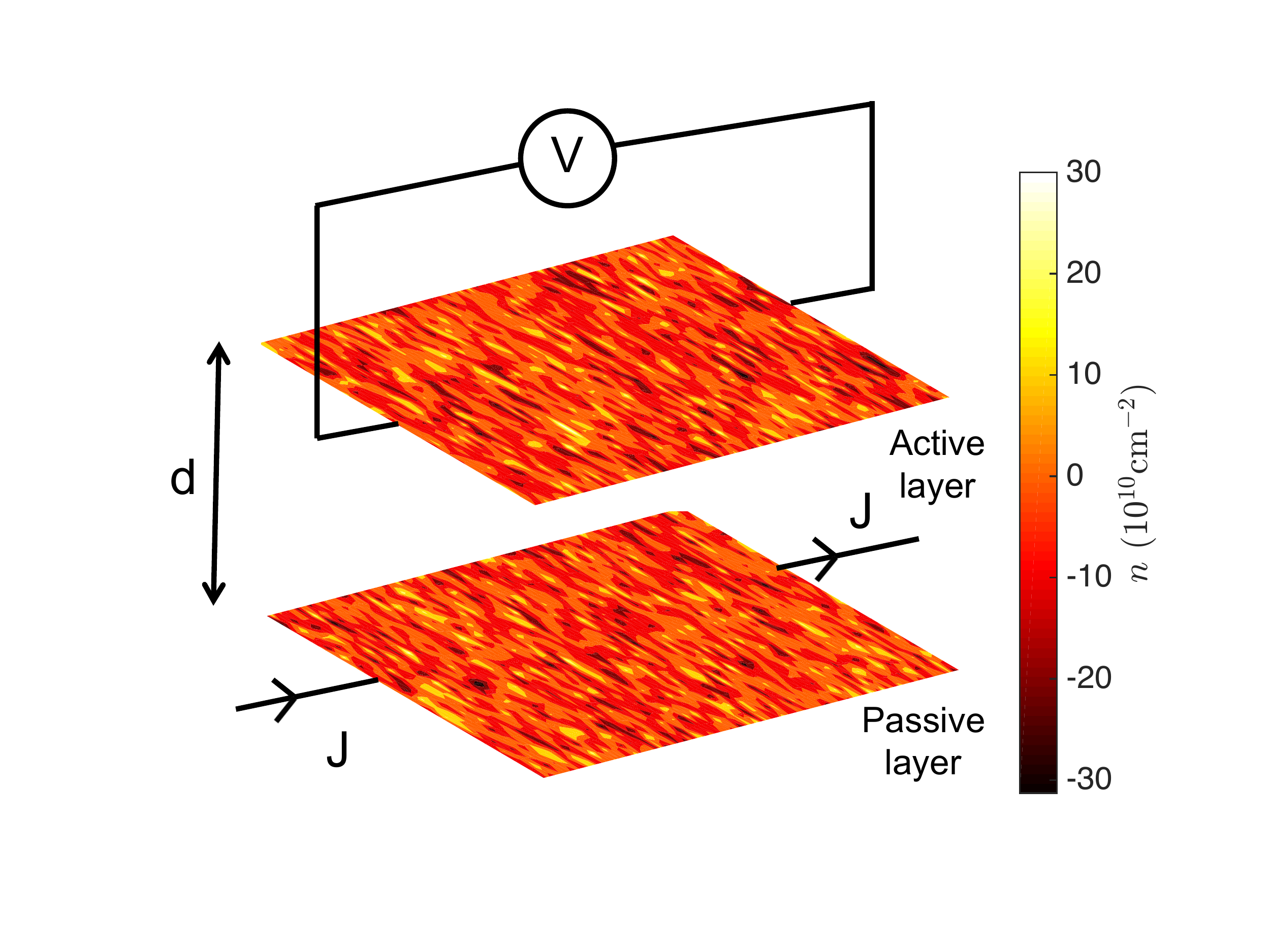}  
		\end{array}$
	\end{center}
	\caption{(color online) Schematic of Coulomb drag between two inhomogeneous sheets of two-dimensional material.}
	\label{inhomg-drag}
\end{figure}

The effects of electron-electron interactions in transport measurements are usually a small correction to predictions from models of non-interacting electrons. Coulomb drag is special because it is identically zero unless interactions are present \cite{narozhny_coulomb_2016}, making it an ideal experimental probe of electron-electron interactions \cite{hu_electron-electron_1996}. A typical experiment measuring drag involves driving a current in one (active) layer and measuring the induced potential drop in a physically separated (passive) layer caused by the Coulomb force  as shown in Fig. \ref{inhomg-drag}. 
The corresponding induced electric field is then divided by the current density in the active layer to yield the drag resistivity. Studies of this effect now have a history of almost thirty years. The first experiments \cite{solomon_new_1989, gramila_mutual_1991,sivan_coupled_1992} were performed using double layer two-dimensional electronic gases (i.e.~GaAlAs heterostructures), followed by a series of associated theoretical works \cite{jauho_coulomb_1993, flensberg_coulomb_1994, kamenev_coulomb_1995, flensberg_linear-response_1995}. A subject of special interest in drag studies is the formation and detection of interlayer exciton condensates \cite{nandi_exciton_2012} which hold the potential for application in low power electronics (see Ref.~\cite{banerjee_bilayer_2009} and references therein). Drag measurements are the standard method for detecting exciton condensation since it shows strong signatures in the drag resistivity \cite{mink_probing_2012, efimkin_anomalous_2016}. 

Recent years have seen a sustained experimental effort to understand Coulomb drag in two dimensional materials such as graphene \cite{kim_coulomb_2011,kim_coulomb_2012,gorbachev_strong_2012,titov_giant_2013} and its bilayer \cite{li_excitonic_2016,lee_giant_2016,li_negative_2016,liu_quantum_2016} due to their high level of tunability and the ability to reach unprecedentedly small separations between the layers while still keeping them electrically isolated, both of which are favourable to the formation of exciton condensates. These two-dimensional layers however also come with a drawback. 
Due to their two-dimensional nature, they tend to possess charge density fluctations \cite{martin_observation_2008} that arise either due to external charged impurities \cite{adam_self-consistent_2007} or corrugations in the topography of the sheets \cite{gibertini_electron-hole_2012}, as shown in Fig. \ref{inhomg-drag}. 
As such inhomogeneity is known to play a role in the single layer carrier transport of two-dimensional materials \cite{adam_self-consistent_2007}, it is natural to expect that they will also play a role in double layer drag transport. Up till now however, there has not been a method for  systematically including the inhomogeneity of the layers in calculations of drag resistivity. 
The situation is very different when it comes to single layer resistivity, for which there exists a well-known formalism that successfully describes charge and heat transport in an inhomogeneous layer known as effective medium theory (EMT) \cite{landauer_electrical_1978,stroud_effective_1998,choy_effective_2016}.
In this work, we close the gap by generalizing EMT for the first time to the case of Coulomb drag between two inhomogeneous sheets and demonstrate the importance of inhomogeneity in drag transport by applying the resulting formalism to two examples. First, we show that in the case of interlayer exciton condensation \cite{seamons_coulomb_2009,nandi_exciton_2012}, where a divergent drag resistivity is expected at zero temperature \cite{vignale_drag_1996}, charge density fluctuations yield a finite value determined by the amplitude of spatial density fluctuations. Next, we apply the drag EMT formalism to drag between two inhomogeneous graphene sheets. 
The standard homogeneous theory predicts drag resistivity peaks that decrease as temperature increases, in contradiction with experiment by Gorbachev {\itshape et al.} \cite{gorbachev_strong_2012} where the opposite is seen. We show that upon inclusion of density inhomogeneity, the drag resistivity peaks increase with temperature within the range of experiment, thus resolving the contradiction. Gorbachev {\itshape et al.} also report measuring anomalously straight drag resistivity isolevels whereas standard homogeneous theory predicts curved ones. 
We demonstrate that these straight isolevels are in fact caused by the presence of charge density fluctuations. 
Lastly, there is an ongoing controversy surrounding the nature of correlations between the layers' fluctuations. As we discuss in detail later, there exist arguments that they are correlated \cite{song_energy-driven_2012}, anti-correlated \cite{gorbachev_strong_2012} or simply uncorrelated. We demonstrate using drag EMT that it is possible to deduce the nature of the correlations by measuring drag resistivity along different lines in the two-layer density parameter space.

The plan of this paper is as follows. Sec. II. presents the derivation of Coulomb drag EMT, of which several example applications will be given in the next two sections. Sec. III investigates excitonic drag in the presence of density fluctuations and Sec. IV studies the impact of these fluctuations on drag between graphene sheets. Sec. V concludes with a discussion of this work and the problems that may be pursued in future based on it.

\section{II. Coulomb Drag Effective Medium Theory} \label{EMT-derivation}

\begin{figure}[h!]
	\begin{center}$
		\begin{array}{c}
		\includegraphics[trim=0cm 3cm 0cm 3cm,clip=true,height=! ,width=9cm] {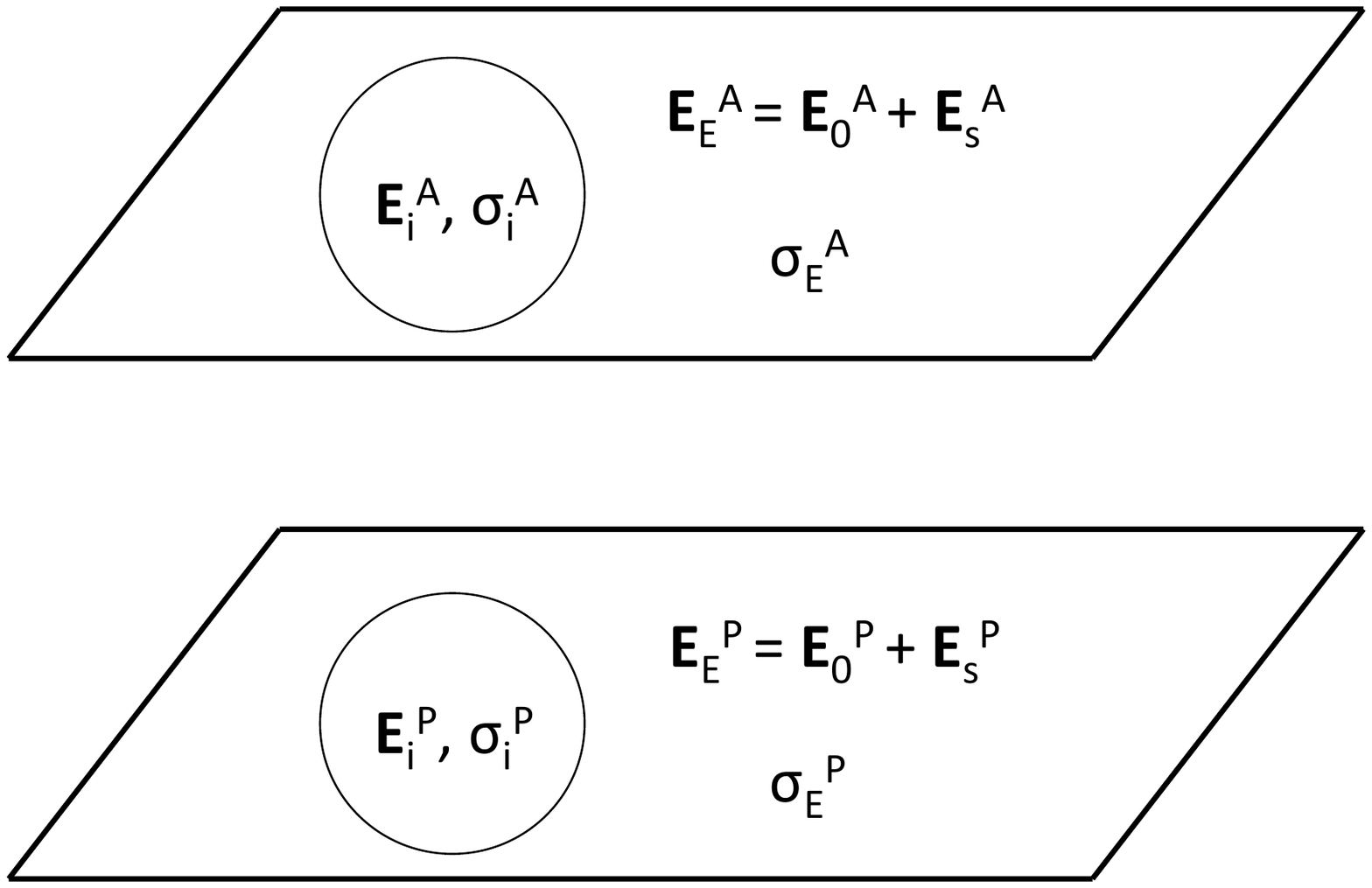}  
		\end{array}$
	\end{center}
	\caption{(color online) The $i$th puddles of the active and passive layers are each embedded inside their own effective media.}
	\label{layers}
\end{figure}

We consider the standard drag setup -- two parallel 2D sheets of identical size separated by some finite distance, with a current flowing through the active layer while the passive layer remains an open circuit. To model the presence of inhomogeneity, we assume that the active and passive layers are each made up of $N$ patches (commonly referred to as `puddles') each with its own conductivity, $\sigma^\act_{i}$ and $\sigma^\pas_{i}$ respectively, where $i=1, \cdots, N$. We assume that the puddles in both layers are circles of radius $a$, with the $i$th puddle of the active layer lying exactly atop the $i$th puddle of the passive, as in Fig.~\ref{layers}. Both puddles are assumed to be of equal area. This assumption is of general applicability because we are allowed to define as many circular patches and make them as small as we wish. 
We do not make any further assumptions about the nature of these puddles and our derivation is applicable regardless of whether the puddles are correlated, anti-correlated or uncorrelated. 

Unlike the standard single layer case (see Appendix A), there are three effective conductivities to be determined. These are the effective in-plane conductivities of the active and passive layers, and the effective drag conductivity between the layers. We denote them by $\sigma^\act_\eff, \sigma^\pas_\eff$ and $\sigma^\drg_\eff$ respectively. 
Our final result is the set of equations (see Eqs.~(\ref{sigmaEMT-drag-cont}) and (\ref{sigmaEMT-mono})) that are solved to yield the effective conductivities.

We summarize the steps of our derivation before delving into the details. First, we take an arbitrary $i$th pair of puddles, one from each layer, and  embed each one inside its own homogeneous effective medium of conductivity $\sigma^\act_\eff$ and $\sigma^\pas_\eff$ respectively as shown in Fig.~\ref{layers}, where each effective medium has within it the uniform effective field (excluding the field caused by the puddle) denoted by $\vec{E}^\act_{0}$ and $\vec{E}^\pas_{0}$ respectively and the drag conductivity between the two effective media is denoted $\sigma^\drg_\eff$. Next, we determine the electric fields inside the puddles, $\vec{E}^\act_{i}$ and $\vec{E}^\pas_{i}$. Finally, we substitute these into the EMT self-consistency equations, given by
\begin{equation}
\sum_{i} f_{i} \vec{E}^\act_{i} = \vec{E}^\act_{0}, \label{EMT-act}
\end{equation}
and
\begin{equation}
\sum_{i} f_{i} \vec{E}^\pas_{i} = \vec{E}^\pas_{0} \label{EMT-pass}
\end{equation}
where $f_{i}$ refers to the areal fraction of the $i$th patch relative to the whole layer to obtain Eqs. (\ref{sigmaEMT-drag}), (\ref{sigmaEMT-active}) and (\ref{sigmaEMT-passive}). Taking the continuum limit, we obtain the final results of Eqs.~(\ref{sigmaEMT-drag-cont}) and (\ref{sigmaEMT-mono}).

We now begin the detailed derivation. Our first task is to solve for the field inside the $i$th (where $i$ is arbitrary) puddle that has been embedded in the effective medium as described above.
We assume that the puddles are regions of uniform 2D polarization in a direction parallel to the effective electric field of the layer. These polarizations are denoted by $\vec{M}^\act = M^\act \vec{e}_x$ and $\vec{M}^\pas = M^\pas \vec{e}_x$ for the active and passive layers respectively. In the active effective medium, we have 
\begin{equation}
U^{\act}_{\eff}(r,\theta) = -E^{\act}_{0}r \cos(\theta) + \frac{a^{2}}{2\epsilon_{0}} \frac{M^{\act}}{r}\cos(\theta) \label{UAE}
\end{equation}
and
\begin{equation}
\vec{E}^{\act}_{\eff}(r,\theta) = E^{\act}_{0} \vec{e}_{x} + \frac{a^{2}}{2\epsilon_{0}} \frac{1}{r^{2}} \left[ 2 (\vec{M}^{\act} \cdot \vec{e}_{r})\vec{e}_{r} - \vec{M}^{\act} \right]. \label{EAE}
\end{equation}
Note that we have chosen a radial coordinate system with its origin at the center of the two concentric circular puddles. Inside the $i$th puddle of the active effective medium, we guess that the field is simply proportional to the effective field 
\begin{equation}
U^{\act}_{i}(r,\theta) = -C^{\act} E^{\act}_{0} r \cos(\theta) \label{UAi}
\end{equation}
and
\begin{equation}
\vec{E}^{\act}_{i}(r,\theta) = C^{\act} E^{\act}_{0} \vec{e}_{x}, \label{EAi}
\end{equation}
where $C^{\act}$ is an unknown constant to be determined. 
The exact same considerations apply for the passive layer. That is, the previous four equations with all `$\mathrm{A}$' superscripts replaced by `$\mathrm{P}$'s describe the passive layer.

We thus have a total of four unknowns, $M^\act$, $M^\pas$, $C^\act$ and $C^\pas$, the last two of which give us the fields within the $i$th puddle of each layer. We solve for these unknowns by making use of boundary conditions. First, the potentials must be continuous at the boundaries of the puddle in each layer. That is,
\begin{equation}
U_{E}^{\act} (r=a,\theta) = U_{i}^{\act} (r=a,\theta) \label{UcontA}
\end{equation}
and
\begin{equation}
U_{E}^{\pas} (r=a,\theta) = U_{i}^{\pas} (r=a,\theta) \label{UcontP}
\end{equation}
for all $\theta$, with the explicit forms of the potentials as given in the previous paragraph.
Second, the radial current density must also be continuous at these boundaries. 
The current densities in the active and passive effective media $\vec{j}_\eff^\act$ and $\vec{j}_\eff^\pas$ are related to the electric fields $\vec{E}_\eff^\act$ and $\vec{E}_\eff^\pas$ by the matrix equation
\begin{equation}
\left( \begin{array}{c} 
\vec{j}_\eff^\act \\ 
0 \end{array} \right) = \left( \begin{array}{cc} 
\sigma_\eff^\act & \sigma_\eff^\drg \\ 
\sigma_\eff^\drg & \sigma_\eff^\pas \end{array} \right) \left( \begin{array}{c} 
\vec{E}_\eff^\act \\ 
\vec{E}_\eff^\pas \end{array} \right), \label{matrixrel1}
\end{equation}
where $\vec{j}_\eff^\pas = 0$ because we consider the situation in which the passive layer is in an open-circuit configuration. Note that $\vec{E}_\eff^\act$ points in the direction of the driving current in the active layer.
Within the puddles, we have the similar relation
\begin{equation}
\left( \begin{array}{c} 
\vec{j}_{i}^\act \\ 
0 \end{array} \right) = \left( \begin{array}{cc} 
\sigma_{i}^\act & \sigma_{i}^\drg \\ 
\sigma_{i}^\drg & \sigma_{i}^\pas \end{array} \right) \left( \begin{array}{c} 
\vec{E}_{i}^\act \\ 
\vec{E}_{i}^\pas \end{array} \right). \label{matrixrel2}
\end{equation}
The electric fields $\vec{E}_\eff^\act$ and $\vec{E}_\eff^\pas$ are either parallel or anti-parallel on physical grounds, since the latter is caused by the former through the drag effect. We choose our axes so that both fields are along the $x$-axis. Explicitly, $\vec{E}_\eff^\act = E_\eff^\act \vec{e}_x$ and $\vec{E}_\eff^\pas = E_\eff^\pas \vec{e}_x$.
Equations (\ref{matrixrel1}) and (\ref{matrixrel2}) together with the requirement of continuous radial current density yield
\begin{equation}
\sigma_{\eff}^{\act} E_{\eff,r}^{\act} + \sigma_{\eff}^{\drg} E_{\eff,r}^{\pas} = \sigma_{i}^{\act} E_{i,r}^{\act} + \sigma_{i}^{\drg} E_{i,r}^{\pas} \label{JcontA}
\end{equation}
and
\begin{equation}
\sigma_{\eff}^{\drg} E_{\eff,r}^{\act} + \sigma_{\eff}^{\pas} E_{\eff,r}^{\pas} = \sigma_{i}^{\drg} E_{i,r}^{\act} + \sigma_{i}^{\pas} E_{i,r}^{\pas}, \label{JcontP}
\end{equation}
where the subscript $r$ denotes the radial component. Substituting the potentials and fields in Eqs. (\ref{UAE}) to (\ref{EAi}) and their counterparts for the passive layer into Eqs. (\ref{UcontA}), (\ref{UcontP}), (\ref{JcontA}) and (\ref{JcontP}) yields four equations for the four unknowns mentioned. Note that the $\theta$ dependence drops out of the problem, leaving behind only the magnitudes of the various vectors. We solve the simultaneous equations for $C^\act$ and $C^\pas$ to find
\begin{widetext}
\begin{equation}
	C^\act= \left(1-\frac{2\left( \sigma^\drg_{i} \sigma^\pas_\eff  - \sigma^\drg_\eff \sigma^\pas_{i} \right)\frac{E^\pas_0}{E^\act_0}
		+ (\sigma^\act_{i}-\sigma^\act_\eff)(\sigma^\pas_\eff+\sigma^\pas_{i})-((\sigma^\drg_{i})^{2}-(\sigma^\drg_\eff)^{2})}{(\sigma^\act_\eff+\sigma^\act_{i})(\sigma^\pas_\eff+\sigma^\pas_{i})-(\sigma^\drg_\eff+\sigma^\drg_{i})^{2}} \right) \label{Cact}
	\end{equation}
and
\begin{equation}
	C^\pas= \left(1-\frac{2\left( \sigma^\drg_{i} \sigma^\act_\eff  - \sigma^\drg_\eff \sigma^\act_{i} \right)\frac{E^\act_0}{E^\pas_0}
		+ (\sigma^\pas_{i}-\sigma^\pas_\eff)(\sigma^\act_\eff+\sigma^\act_{i})-((\sigma^\drg_{i})^{2}-(\sigma^\drg_\eff)^{2})}{(\sigma^\act_\eff+\sigma^\act_{i})(\sigma^\pas_\eff+\sigma^\pas_{i})-(\sigma^\drg_\eff+\sigma^\drg_{i})^{2}} \right). \label{Cpas}
	\end{equation}
	\end{widetext}
We have thus solved for the electric fields within the $i$th pair of puddles. In order for these fields to be combined with the self-consistency equations in a useful manner however, we must first be able to write the electric field inside each puddle as a function of only its own layer's effective medium field (i.e., instead of being a function of the effective medium fields of both layers). We achieve this by requiring that just given the two effective medium layers without puddles (i.e., Fig. \ref{layers} with the puddles taken out), there will be no current flow in the passive layer. Physically, this means that $E^\act_{0}$ and $E^\pas_{0}$ represent uniform fields that effectively model the spatially fluctuating fields in the two layers. The expression for this condition is given by
\begin{equation}
\frac{E^\act_{0}}{E^\pas_{0}} = - \frac{\sigma^\pas_\eff}{ \sigma^\drg_\eff}. \label{zerojE}
\end{equation}
Substituting Eqs. (\ref{Cact}) and (\ref{Cpas}) into Eqs. (\ref{EAi}) and its passive layer counterpart respectively and making use of Eq.~(\ref{zerojE}), we obtain the intra-puddle field in the active (passive) layer as a function of only the active (passive) layer's effective medium electric field. Explicitly, we obtain for the active layer puddle
\begin{widetext}
\begin{equation}
	E^\act_{i}= \left(1-\frac{2\left( \sigma^\drg_\eff \sigma^\pas_{i}  - \sigma^\drg_{i} \sigma^\pas_\eff \right)\frac{\sigma^\drg_\eff}{\sigma^\pas_\eff}+(\sigma^\act_{i}-\sigma^\act_\eff)(\sigma^\pas_\eff+\sigma^\pas_{i})-
		((\sigma^\drg_{i})^{2}-(\sigma^\drg_\eff)^{2})}
	{(\sigma^\act_\eff+\sigma^\act_{i})(\sigma^\pas_\eff+\sigma^\pas_{i})-(\sigma^\drg_\eff+\sigma^\drg_{i})^{2}} \right) E^\act_{0}
\end{equation}
and 
	\begin{equation}
	E^\pas_{i}= \left(1-\frac{2\left( \sigma^\drg_\eff \sigma^\act_{i} - \sigma^\drg_{i} \sigma^\act_\eff \right)\frac{\sigma^\pas_\eff}{\sigma^\drg_\eff}
		+ (\sigma^\pas_{i}-\sigma^\pas_\eff)(\sigma^\act_\eff+\sigma^\act_{i})-((\sigma^\drg_{i})^{2}-(\sigma^\drg_\eff)^{2})}{(\sigma^\act_\eff+\sigma^\act_{i})(\sigma^\pas_\eff+\sigma^\pas_{i})-(\sigma^\drg_\eff+\sigma^\drg_{i})^{2}} \right) E^\pas_{0}
	\end{equation}
for the passive layer puddle.
\end{widetext}

Finally, we substitute these into the self-consistency equations (\ref{EMT-act}) and (\ref{EMT-pass}) and make the approximation of setting all terms quadratic in drag conductivities to zero since drag conductivities are typically much smaller than in-plane conductivities. This yields 
\begin{equation}
\sum_{i} f_{i} \cdot \frac{\sigma^\act_{i}-\sigma^\act_\eff} 
{\sigma^\act_\eff+\sigma^\act_{i}} = 0. \label{sigmaEMT-active}
\end{equation}
and
\begin{equation}
\sum_{i} f_{i} \cdot \frac{2\left( \sigma^\drg_\eff \sigma^\act_{i} - \sigma^\drg_{i} \sigma^\act_\eff \right)\frac{\sigma^\pas_\eff}{\sigma^\drg_\eff}
}{(\sigma^\act_\eff+\sigma^\act_{i})(\sigma^\pas_\eff+\sigma^\pas_{i})} + 
\sum_{i} f_{i} \cdot \frac{\sigma^\pas_{i}-\sigma^\pas_\eff}{\sigma^\pas_\eff+\sigma^\pas_{i}} =0, \label{2terms}
\end{equation}
where we have made use of the fact that $\sum_{i} f_{i} = 1$. Equation (\ref{sigmaEMT-active}) is the well-known discrete single layer EMT equation applied to the active layer. Equation (\ref{2terms}) is more complicated and comprises two terms summing to zero. Since the two unknowns $\sigma^\drg_\eff$ and $\sigma^\pas_\eff$ cannot be determined by this single equation, we require another condition. Since the drag conductivity is very small compared to the in-plane conductivity within either layer, we may obtain this condition by approximating that the interlayer interaction has a negligible effect on the passive layer conductivity. This then implies that the standard single layer EMT equation (cf.~(\ref{sigmaEMT-active}) ) applies to the passive layer, and the two terms of Eq.~(\ref{2terms}) must both be individually zero. Equating the first term to zero and solving for $\sigma^\drg_\eff$ yields
\begin{equation}
\sigma^\drg_\eff = \sigma^\act_\eff \frac{\sum_{i} f_{i} \cdot \frac{ \sigma^\drg_{i} }
{(\sigma^\act_\eff+\sigma^\act_{i})(\sigma^\pas_\eff+\sigma^\pas_{i})}} {\sum_{i} f_{i} \cdot \frac{ \sigma^\act_{i} }
{(\sigma^\act_\eff+\sigma^\act_{i})(\sigma^\pas_\eff+\sigma^\pas_{i})} } \label{sigmaEMT-drag},
\end{equation} 
while setting the second term to zero yields
\begin{equation}
\sum_{i} f_{i} \cdot \frac{\sigma^\pas_{i}-\sigma^\pas_\eff}{\sigma^\pas_\eff+\sigma^\pas_{i}} =0 \label{sigmaEMT-passive}.
\end{equation}
The former is a newly derived discrete EMT drag equation while the latter is just the already-known discrete single layer EMT equation applied to the passive layer. Note that in the case of there being only two puddles in each layer (relevant only at double charge neutrality) so that $f_{i}= \frac{1}{2}$, $i={1,2}$, we recover the results of Ref.~\cite{apalkov_effective_2005} which considered drag between inhomogeneous layers each consisting of only two areal components. Generalizing Eq. (\ref{sigmaEMT-drag}) to the continuum limit, we obtain the EMT drag conductivity equation
\begin{widetext}
\begin{equation}
\sigma_{\drg}^{\scr{E}}=  \sigma_{\scr{A}}^{\scr{E}} \frac{\int^{\infty}_{-\infty} dn_{\act}'  \int^{\infty}_{-\infty} dn_{\pas}' P_{\mathrm{bi}}(n_{\act}',n_{\pas}') \cdot \left[ \frac{\sigma_{\scr{D}}(n_{\scr{A}}',n_{\scr{P}}') }{(\sigma_{\scr{A}}^{\scr{E}}+\sigma_{\scr{A}}(n_{\scr{A}}'))(\sigma_{\scr{P}}^{\scr{E}}+\sigma_{\scr{P}}(n_{\scr{P}}') )} \right] }{\int^{\infty}_{-\infty} dn_{\act}'  \int^{\infty}_{-\infty} dn_{\pas}' P_{\mathrm{bi}}(n_{\act}',n_{\pas}') \cdot  \left[ \frac{  \sigma_{\scr{A}}(n_{\scr{A}}') }{(\sigma_{\scr{A}}^{\scr{E}}+\sigma_{\scr{A}}(n_{\scr{A}}'))(\sigma_{\scr{P}}^{\scr{E}}+\sigma_{\scr{P}}(n_{\scr{P}}') )}\right]},\label{sigmaEMT-drag-cont}
\end{equation}
\end{widetext}
where $n_{\scr{A}}'$ and $n_{\scr{P}}'$ denote the charge densities in the active and passive layers and $\sigma_{\drg}^{\eff}$ is the effective medium theory averaged drag conductivity obtained by solving the equation. $P_{\mathrm{bi}}(n_{\act}',n_{\pas}')$ is the joint probability distribution of finding two points on the layers with one point lying directly above the other having charge densities $(n_{\act}',n_{\pas}')$.
Doing the same for Eqs. (\ref{sigmaEMT-active}) and (\ref{sigmaEMT-passive}) yields
\begin{equation}
\int^{\infty}_{-\infty} dn_{i}' P_{\mathrm{mono}}(n_{i}') \frac{\sigma_{i}(n_{i}') - \sigma_{i}^{\eff}}{\sigma_{i}(n_{i}') + \sigma_i^{\eff}} = 0, \label{sigmaEMT-mono}
\end{equation}
where $i= \mathrm{A}, \mathrm{P}$ denotes the layer index and $\sigma_{i}(n_{i}')$ is the homogeneous conductivity of layer $i$ at uniform density $n_{i}'$. $P_{\mathrm{mono}}(n_{i}')$ is the single layer probability density of finding a point on layer $i$ with charge density $n_{i}'$.

Equation (\ref{sigmaEMT-drag-cont}) is the main result of this work and represents the first generalization of EMT to the drag problem. We emphasize that it is general and applies to drag between any two sheets of two-dimensional material. The drag resistivity is given by solving Eqs. (\ref{sigmaEMT-drag-cont}) and (\ref{sigmaEMT-mono}) for the three conductivities $\sigma_{\drg}^{\scr{E}}$,  $\sigma^\act_\eff$, $\sigma^\pas_\eff$ and inserting them into 
\begin{equation}
\rho_{\drg}^{\scr{E}} = - \frac{\sigma_{\drg}^{\scr{E}}}{\sigma^\act_\eff \sigma^\pas_\eff - (\sigma_{\drg}^{\scr{E}})^{2}}.  \label{rhoD}
\end{equation}
The standard homogeneous theory \cite{flensberg_linear-response_1995,kamenev_coulomb_1995,narozhny_coulomb_2012} is recovered in the limit of $ n_{\mathrm{rms}}^{(\act,\pas)} \rightarrow 0$.

To perform actual calculations, one must choose specific probability distributions for $P_{\mathrm{mono}}$ and $P_{\mathrm{bi}}$. We choose for the former the usual monovariate Gaussian distribution, 
\begin{eqnarray}
P_{\mathrm{mono}}(n_{i}') & \equiv & P_{\mathrm{mono}}(n_{i}';n_{i},n_{\mathrm{rms}}^{(i)}) \\ \nonumber
						 & = & \frac{1}{\sqrt{2 \pi} n_{\mathrm{rms}}^{(i)} } \exp\left(- \frac{(n_{i}'-n_{i})^{2}}{2 (n_{\mathrm{rms}}^{(i)})^{2} } \right), \label{gaussian}
\end{eqnarray}
where $n_{i}$ without the prime superscript denotes the average charge density of layer $i$ set by the external gate voltage. $n_{\mathrm{rms}}^{(i)}$ is the root mean square density fluctuation about the average caused by charged impurities and quantifies the strength of inhomogeneity in the sample.

We model the double layer distribution using the bivariate normal probability distribution
\begin{widetext}
	\begin{eqnarray}
	P_{\mathrm{bi}}(n_{\act}',n_{\pas}') & \equiv & P_{\mathrm{bi}}(n_{\act}',n_{\pas}';n_{\act},n_{\pas},n_{\mathrm{rms}}^{\act},n_{\mathrm{rms}}^{\pas},\eta) \\ \nonumber
	&=&\frac{1}{2\pi n^\act_{\mathrm{rms}} n^\pas_{\mathrm{rms}} \sqrt{1-\eta^{2}}} \exp \left( -\frac{1}{2(1-\eta^{2})} \left[ \frac{(n_{\act}' - n_{\act})^{2} }{(n^\act_{\mathrm{rms}})^{2}} + \frac{(n_{\pas}' - n_{\pas})^{2} }{(n^\pas_{\mathrm{rms}})^{2}} - \frac{2\eta(n_{\act}' - n_{\act})(n_{\pas}' - n_{\pas})}{n^\act_{\mathrm{rms}}n^\pas_{\mathrm{rms}}} \right] \right), \label{bivariate-norm-distrib}
	\end{eqnarray} 
\end{widetext}
where the interlayer correlation coefficient $\eta$ quantifies the charge density fluctuations between the two layers. A value of $\eta=1$ ($-1$) corresponds to perfectly correlated (anti-correlated) charge density fluctuations within the two layers, while a value of $\eta=0$ corresponds to uncorrelated fluctuations. Mathematically, $\eta$ is defined by
\begin{equation}
\eta \equiv \frac{\langle (n_{\act}' - n_{\act}) (n_{\pas}' - n_{\pas})\rangle}{n^\act_{\mathrm{rms}}n^\pas_{\mathrm{rms}}},
\end{equation} 
where the angular brackets refer to averaging over the areas of the two layers.  
Static charged impurities in the surroundings of two sheets held close together can lead to correlated fluctuations, whereas random strain in the sheets together with strong Coulomb coupling between them can lead to anti-correlated fluctuations.
If the separation between layers is fairly large, the fluctuations will tend to be uncorrelated due to each sheet seeing a potential of different origin. 
All these scenarios may be modeled in Eq.~(\ref{bivariate-norm-distrib}) by choosing the value of $\eta$ accordingly. In an earlier version of this work, we claimed that Onsager reciprocity relation is violated for $\eta \neq 0$ but this has since been found to be due to a numerical error. We show in Appendix B a proof that Onsager reciprocity is obeyed in the drag EMT formalism.

Before moving on to applications, we discuss the physical conditions under which inhomogeneities strongly affect transport and the drag EMT must be applied.
There are three energy scales that need to be considered. 
First, the temperature of the system $k_{B}T$ is important. 
Second, we introduce the typical layer Fermi energy ${E}_{F}(\bar{n})$ where $\bar{n} \equiv \sqrt{n_{\act} n_{\pas}}$ is the typical average layer density. 
Lastly, we define an inhomogeneity energy scale $E_{F}(n^{*})$ where $n^{*} \equiv \sqrt{n_{\mathrm{rms}}^{\act} n_{\mathrm{rms}}^{\pas}}$ is the typical root mean square fluctuation of density in the layers.
Generally speaking, the impact of fluctuations (i.e. the amount by which drag changes after including density fluctuations) is strong when 
\begin{equation}
E_{F}^{*} \sim \mathrm{max}(\bar{E}_{F},k_{B}T) \label{EMT-impt}
\end{equation}
is satisfied.
The reason is as follows. The drag EMT essentially yields an average of the homogeneous $\rho_{\drg}$ over a region of density space centered at average densities $(n_{\act},n_{\pas})$ with an area on the order of $(n^{*})^{2}$, with the exact shape and orientation of the averaging region determined by $\eta$ and the relative magnitudes of $n_{\mathrm{rms}}^{\act}$ and $n_{\mathrm{rms}}^{\pas}$. 
Such a coarse-graining procedure makes a big difference when performed over regions in which the function being averaged possesses turning points.
In the case of drag resistivity, these occur at the double neutrality point, and in the regions $|\bar{E}_{F}| \sim k_{B} T$ where the finite density drag peaks occur, and they will be inside the region of averaging when Eq. (\ref{EMT-impt}) is true. 
Inhomogeneity is negligible if $E_{F}^{*} \ll \mathrm{max}(\bar{E}_{F},k_{B}T)$ because the averaging encompasses a region over which $\rho_{\drg}$ does not change much. Lastly, if $E_{F}^{*} \gg \mathrm{max}(\bar{E}_{F},k_{B}T)$, then $\rho_{\drg}^{\eff}$ will be approximately zero everwhere since the averaging window includes the high density regions where drag has already gone to zero for all intents and purposes.

\section{III. IMPACT OF INHOMOGENEITY ON EXCITONIC DRAG} 

\begin{figure}[h!]
	\begin{center}$
		\begin{array}{c}
		\includegraphics[trim=0cm 3cm 0cm 3cm,clip=true,height=! ,width=9cm] {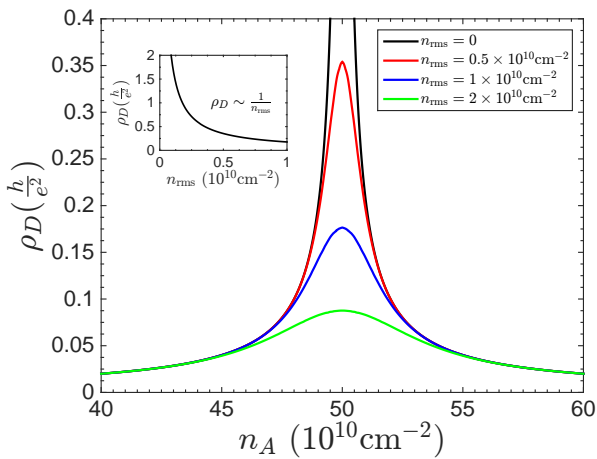}  
		\end{array}$
	\end{center}
	\caption{(color online) Drag resistivity in the regime of exciton condensation as a function of $n_{A}$ for different strengths of charge density inhomogeneity  $n_{\mathrm{rms}}^{(\act)}= n_{\mathrm{rms}}^{(\pas)} \equiv n_{\mathrm{rms}}$. $n_{P}$ is held constant at $-50 \times 10^{10} \mathrm{cm}^{-2}$.  Inset: $\rho_{D}$ at $n_{A} = -n_{P} = 50 \times 10^{10} \mathrm{cm}^{-2}$ as $n_{\mathrm{rms}}$ is varied.}
	\label{exciton-div}
\end{figure}  

Under the right conditions, electrons and holes in the two layers are expected to bind together forming stable bosonic excitons, which can condense into a superfluid exciton condensate (see for instance Refs. \cite{min_room-temperature_2008} and \cite{lozovik_condensation_2012}). In particular, the drag resistivity of an exciton condensate diverges as temperature approaches zero \cite{vignale_drag_1996} because the magnitude of drag conductivity approaches that of the single layer conductivity. We demonstrate that this divergence is suppressed in the presence of charge density inhomogeneity. We model the exciton condensate at zero temperature using the following phenomenological expressions for the various conductivities. The monolayer conductivity is given by
\begin{equation}
\frac{\sigma_{i}}{\sigma_{0}} = A \left|\frac{n_{i}}{n_{0}} \right|^{\alpha} ,\label{sigma-mono-ex}
\end{equation}
and the drag conductivity by
\begin{eqnarray}
\frac{\sigma_{D}}{\sigma_{0}} &=& -A \left(\frac{ \mathrm{min} (n_{\act},n_{\pas}) }{n_{0}}\right)^{\alpha} \left( \frac{1-\mathrm{sgn}(n_{\act}n_{\pas})}{2}\right) \\ \nonumber
 && + 10^{-2}  \left| \frac{n_{\act} n_{\pas} }{n_{0}} \right|^{1/2} \left( \frac{1+\mathrm{sgn}(n_{\act}n_{\pas})}{2}\right).\label{sigmad-ex}
\end{eqnarray} 

In the above, $i= {A,P}$ is a layer index, and $\sigma_{0} = \frac{e^{2}}{h}$, $n_{0}=10^{10} \mathrm{cm}^{-2}$. $A$ and $\alpha$ are phenomenological coefficients that can be given arbitrary values depending on the specific material under consideration. 
We note that the above expressions produce the correct behavior in various limits. 
When the passive layer is in open-circuit configuration and no current flows in it, they lead to equal electric fields in the two layers, as expected from Ref. \cite{vignale_drag_1996}. 
When the passive layer is short-circuited so that there is no electric field across it, they yield equal charge currents in the two layers.

In the absence of inhomogeneity (ie. $n_{\mathrm{rms}}^{(\act,\pas)} =0$), it is clear from Eq. (\ref{rhoD}) that a divergence in $\rho_{D}$ occurs at perfectly matched opposite densities $n_{\act} = -n_{\pas}$. To investigate the effect of density fluctuations on exciton drag, we substitute the above conductivity expressions into the EMT equations (\ref{sigmaEMT-drag-cont}) and (\ref{sigmaEMT-mono}) and use the probability distributions in Eqs. (\ref{gaussian}) and (\ref{bivariate-norm-distrib}) with various values of $n_{\mathrm{rms}}^{(\act,\pas)}$. As shown in Fig. \ref{exciton-div}, density fluctuations suppress the divergence in drag resistivity. Furthermore, the magnitude of drag at perfectly matched densities goes inversely as $n_{\mathrm{rms}}^{(\act,\pas)}$ as shown in the inset. Here, we have used $A=5$, $\alpha=1$ and $\eta=0$ but our numerics suggest that these statements still apply for arbitrary values of $A$ and all positive powers $\alpha$. They also apply regardless of the value of correlation coefficient $\eta$. 
This finding demonstrates that sample inhomogeneity is important when using drag resistivity as a probe of exciton condensation and should be of great interest in the ongoing search for exciton condensation \cite{li_excitonic_2016,liu_quantum_2016,de_cumis_complete_2016}.

\section{IV. DRAG IN GRAPHENE SETUPS}

In this section, we demonstrate that just as in single layer graphene transport, inhomogeneity plays an important role in graphene drag transport and it is necessary to take inhomogeneity into account in order to explain the experimental data in the literature. We begin with a brief review of drag calculation.

\subsection{1. Review of Coulomb drag theory}
Coulomb drag has been studied theoretically in graphene monolayers in several works \cite{tse_theory_2007,peres_coulomb_2011,narozhny_coulomb_2012,carrega_theory_2012,amorim_coulomb_2012}. 
The drag conductivity $\sigma_{\drg}(n_{\act},n_{\pas})$ between two sheets at uniform densities $n_{\act}$ and $n_{\pas}$ respectively can be derived diagramatically \cite{flensberg_linear-response_1995,kamenev_coulomb_1995} as
\begin{eqnarray}
\sigma_{\drg} (n_{\act},n_{\pas}) &=& \frac{1}{16 \pi k_{\scr{B}} T} \int \displaylimits_{-\infty}^{\infty} \frac{d^{2}q}{(2\pi)^2} \int \displaylimits_{-\infty}^{\infty} \frac{d\omega}{\sinh^{2}(\frac{\hbar\omega}{2k_{\scr{B}}T})}  \nonumber \\
&\times & \Gamma ^x _{\act} \left(n_{\act}, \mathbf{q} ,\omega \right) \Gamma ^x _{\pas} \left(n_{\pas},\mathbf{q}, \omega \right) |V(d)|^{2},  \label{sigmaD}
\end{eqnarray}
where $T$ is temperature, and $d$ the interlayer spacer width. $V(d)$ is the dynamically screened interlayer Coulomb interaction \cite{ramezanali_finite-temperature_2009}, and $\Gamma ^x$ refers to the $x$-component of the nonlinear susceptibility in monolayer graphene as given in Ref. \cite{narozhny_coulomb_2012}. In our calculations here, we choose parameters based on the experimental setup in Gorbachev {\itshape et al.} \cite{gorbachev_strong_2012} where both graphene sheets are encapsulated in hexagonal boron nitride (hBN) and separated by a hBN spacer. Further details on these quantities may be found in Appendix C. In all our calculations, we use the following parameters unless otherwise specified- the interlayer spacing is $d=9 \mathrm{nm}$,  distance of the active (passive) layer from the charged impurity plane is $20\mathrm{nm}$ ($10\mathrm{nm}$) and the impurity concentration on the charged impurity plane is $n_{\mathrm{imp}} = 15 \times 10^{10} \mathrm{cm}^{-2}$.

The in-plane conductivity of layer $i$ at uniform density $n_{i}$ is given by 
\begin{equation}
\sigma_{i}(n_{i}) = \frac{e^{2}v_{F}^{2}}{2} \int dE D(E)  \tau_{i}(E) \left( -\frac{\partial f(E) }{\partial E} \right), \label{sigmamono}
\end{equation}
where $i=A,P$, $D(E) = 2 |E| / (\pi \hbar^{2} v_{F}^{2})$ is the density of states and $f(E)$ the Fermi function, given by 
$f(E,\mu_{i}) = \left( \exp(\frac{E-\mu_{i}}{k_{\scr{B}} T}) + 1 \right)^{-1} $
and $\tau_{i}(E)$ is the intralayer transport scattering time. The chemical potential $\mu_{i}$ is determined by $n_{i}$ and $T$ (see Appendix C). 
We assume that electron-charged impurity scattering is the dominant scattering mechanism and neglect all others. 
The previous two equations together with Eq. (\ref{rhoD}) yield drag resistivity at any point in density space $(n_{\act}, n_{\pas})$, assuming completely homogeneous graphene sheets.

\subsection{2. Temperature Dependence of drag resistivity peaks}

\begin{figure}[h!]
	\begin{center}$
		\begin{array}{c}
		\includegraphics[trim=0cm 6.5cm 0cm 6.5cm,clip=true,height=! ,width=8cm] {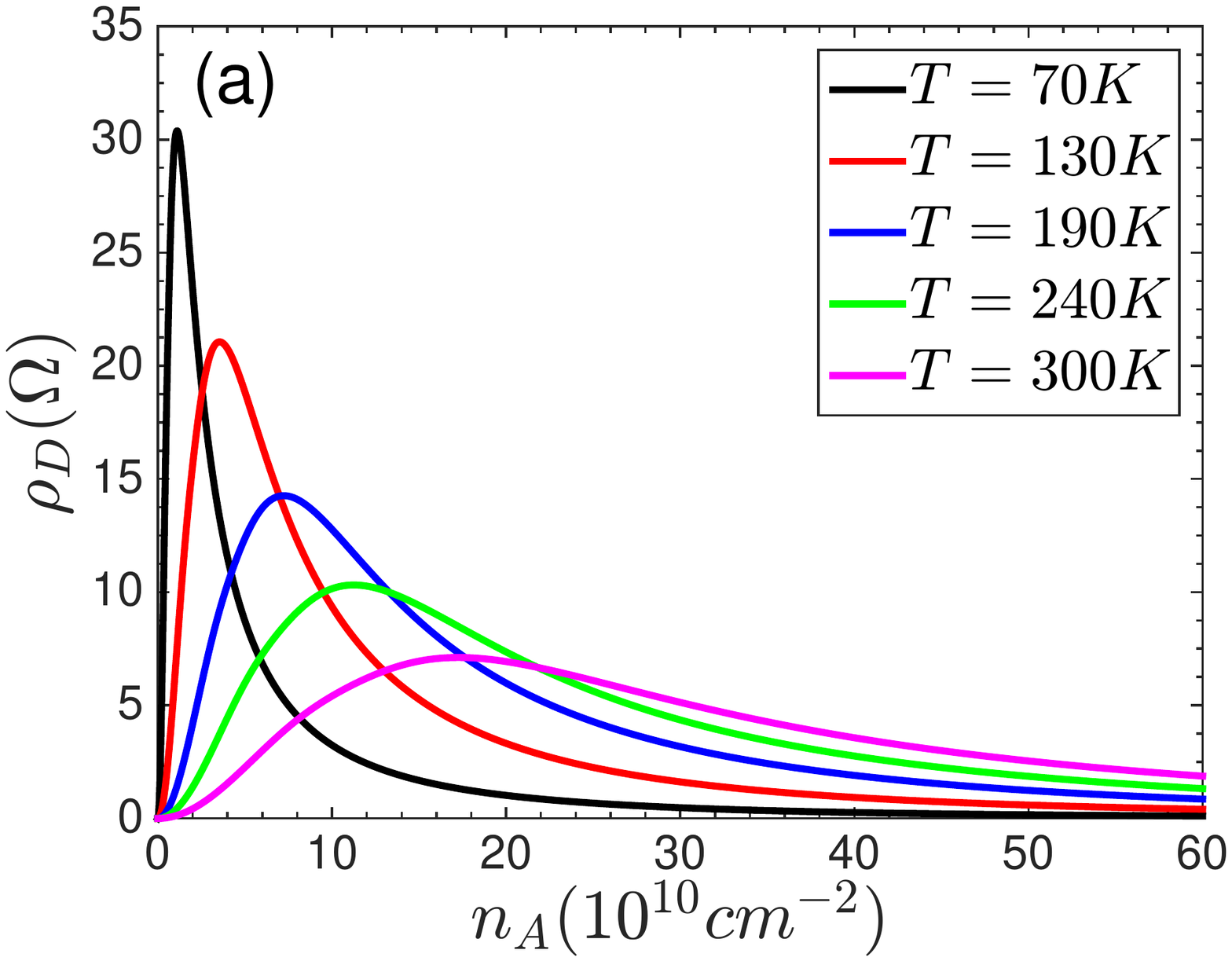}  \\ 
		\includegraphics[trim=0cm 6.5cm 0cm 6.5cm,clip=true,height=! ,width=8cm] {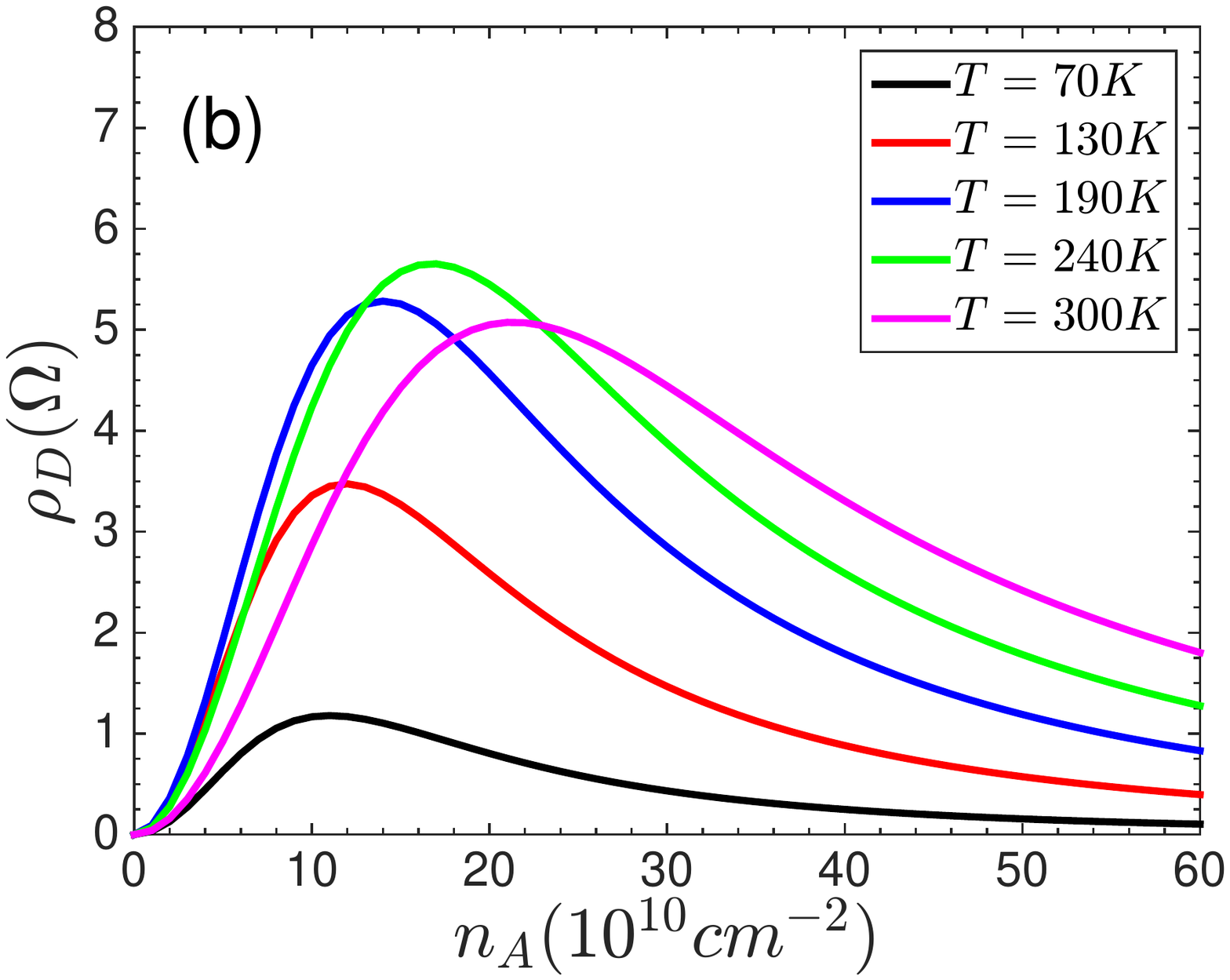}  \\
		\end{array}$
	\end{center}
	\caption{(color online) Re-ordering of drag resistivity peaks by disorder. (a) Drag resistivity in the homogeneous theory. The peaks in $\rho_{D}$ decrease monotonically with temperature in contradiction with experiment. (b) Drag resistivity in the inhomogeneous theory calculated using EMT. Here $n_{\mathrm{rms}}^{(A,P)} = (7,14) \times 10^{10} \mathrm{cm}^{-2}$ with $\eta=0$. The peaks in $\rho_{D}$ display a non-monotonic temperature dependence.}
	\label{T-dep-1}
\end{figure}

Using the above expressions, we calculate drag resistivity along the line of oppositely matched densities $n_{\act} = -n_{\pas}$ and show the results in Fig. \ref{T-dep-1}(a). At each temperature, the drag resistivity follows the standard density dependence. At charge neutrality, it is zero due to electron-hole symmetry. As we move away from charge neutrality, drag increases to a peak (henceforth referred to simply as the `drag peak') and subsequently goes to zero at high density as screening between the layers effectively decouples them. However, a surprise occurs as we tune temperature. The drag peaks decrease as temperature increases, in direct contradiction with the experimental observations of Gorbachev {\itshape et al}. This puzzling contradiction appears to have gone unnoticed in the literature.

The resolution lies in including inhomogeneity in the calculation of drag using EMT. This is done by substituting Eqs. (\ref{sigmaD}) and (\ref{sigmamono}) into the EMT equations (\ref{sigmaEMT-drag-cont}) and (\ref{sigmaEMT-mono}) to obtain the effective conductivities $\sigma^{\drg}_{\eff}$, $\sigma^{\act}_{\eff}$ and $\sigma^{\pas}_{\eff}$, which are then substituted into Eq. (\ref{rhoD}) to obtain the effective drag resistivity. We assume uncorrelated density fluctuations $\eta=0$ in Eq. (\ref{bivariate-norm-distrib}) because interlayer correlations at finite densities away from the double neutrality point $n_{\act},n_{\pas}=0$ are expected to be weak due to screening.
The drag peaks thus obtained by EMT are smaller than their values assuming perfect homogeneity because EMT essentially averages $\rho_{\drg}$ over some region in density space and this can only result in maxima decreasing in height. 
The drag peaks also follow a non-monotonic temperature dependence, increasing with temperature to a peak at some temperature set by the inhomogeneity strength, and decreasing thereafter. 
We choose inhomogeneity strength $n_{\mathrm{rms}}^{(\act,\pas)} = (7,14) \times 10^{10} \mathrm{cm}^{-2}$ so that the highest peak occurs at $T=240K$ (i.e. the highest temperature studied in Gorbachev {\itshape et al.}) and show the results of our calculation in Fig. \ref{T-dep-1}(b). 
At high densities and temperatures such that $\hbar v_{F} \sqrt{\pi n_{\act,\pas}}$ and $k_{B} T$ are much greater than $\hbar v_{F} \sqrt{\pi n_{\mathrm{rms}}^{(\act,\pas)}}$, the homogeneous and EMT drag calculations yield essentially the same values, as expected since the disorder energy scale is now the smallest energy scale of the problem.
As a corollary of this, the decrease in drag peak at temperatures above $240K$ occurs because inhomogeneity becomes unimportant and the homogeneous behavior re-emerges.
The decrease in drag peak above $240K$ is a concrete prediction of our theory that should be testable in existing experimental setups.

\begin{figure}[h!]
	\begin{center}$
		\begin{array}{c}
		\includegraphics[trim=0cm 6.5cm 0cm 6.5cm,clip=true,height=! ,width=8cm] {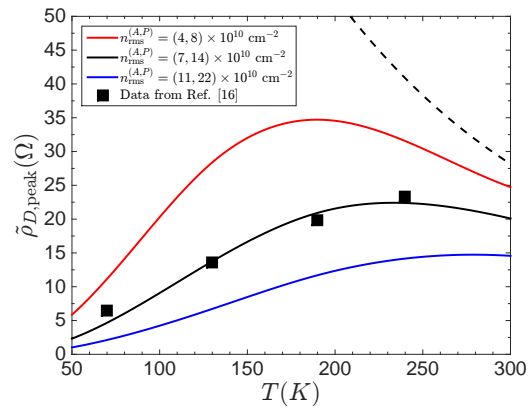}
		\end{array}$
	\end{center}
	\caption{(color online) The behavior of the dressed drag peaks $\tilde{\rho}_{D,\mathrm{peak}}$ for different density fluctuation strengths at $\eta=0$ as a function of temperature. The dashed line shows the homogeneous case.}
	\label{T-dep-2}
\end{figure}

The drag resistivities calculated in Fig. \ref{T-dep-1} are smaller than those of the experiment,  possibly due to enhancement mechanisms such as dielectric inhomogeneity \cite{badalyan_enhancement_2012} and virtual phonons \cite{amorim_coulomb_2012-1} which we have not included in our calculations. We defer a detailed study of these effects to a future work. For now, we take them into account phenomenologically by comparing our theoretically calculated drag resistivity with experiment at high density and temperature 
(where inhomogeneity plays a negligible role) and using the extracted discrepancy factor to scale up our calculated $\rho_{\drg}$ values by hand. 
Comparing the experimental $\rho_{\drg}$ at $n_{\act} = -n_{\pas} = 60 \times 10^{10} \mathrm{cm}^{-2}$ and $T=240K$ (i.e. the largest density and temperature in the data of Gorbachev {\itshape et al.}) with the corresponding $\rho_{\drg}$ calculated using the homogeneous theory, we find a discrepancy of $3.96$. 
A similar discrepancy was also found by Gorbachev {\itshape et al}. 
We thus define a `dressed' drag resistivity $\tilde{\rho}_{\drg}$ as the calculated $\rho_{\drg}$ multiplied by a factor of $3.96$. 
Comparing the dressed EMT drag resistivity peaks at various inhomogeneity strengths with experiment in Fig. \ref{T-dep-2}, we see that the curve for $n_{\mathrm{rms}}^{(A,P)} = (7,14) \times 10^{10} \mathrm{cm}^{-2}$ agrees well with experiment. 
We choose a larger root mean square density fluctuation for one layer because in general one layer is expected to be nearer the impurity plane than the other. 
However, we note that our numerics do not show much difference when the individual layer root mean square fluctuations are changed so long as the total fluctuation $n_{\mathrm{rms}}^{(A)}+n_{\mathrm{rms}}^{(P)}$ is unchanged.

\subsection{3. Drag isolevels in the presence of inhomogeneity}

\begin{figure}[h!]
	\begin{center}$
		\begin{array}{c}
		\includegraphics[trim=0cm 6.5cm 0cm 6.5cm,clip=true,height=! ,width=8cm] {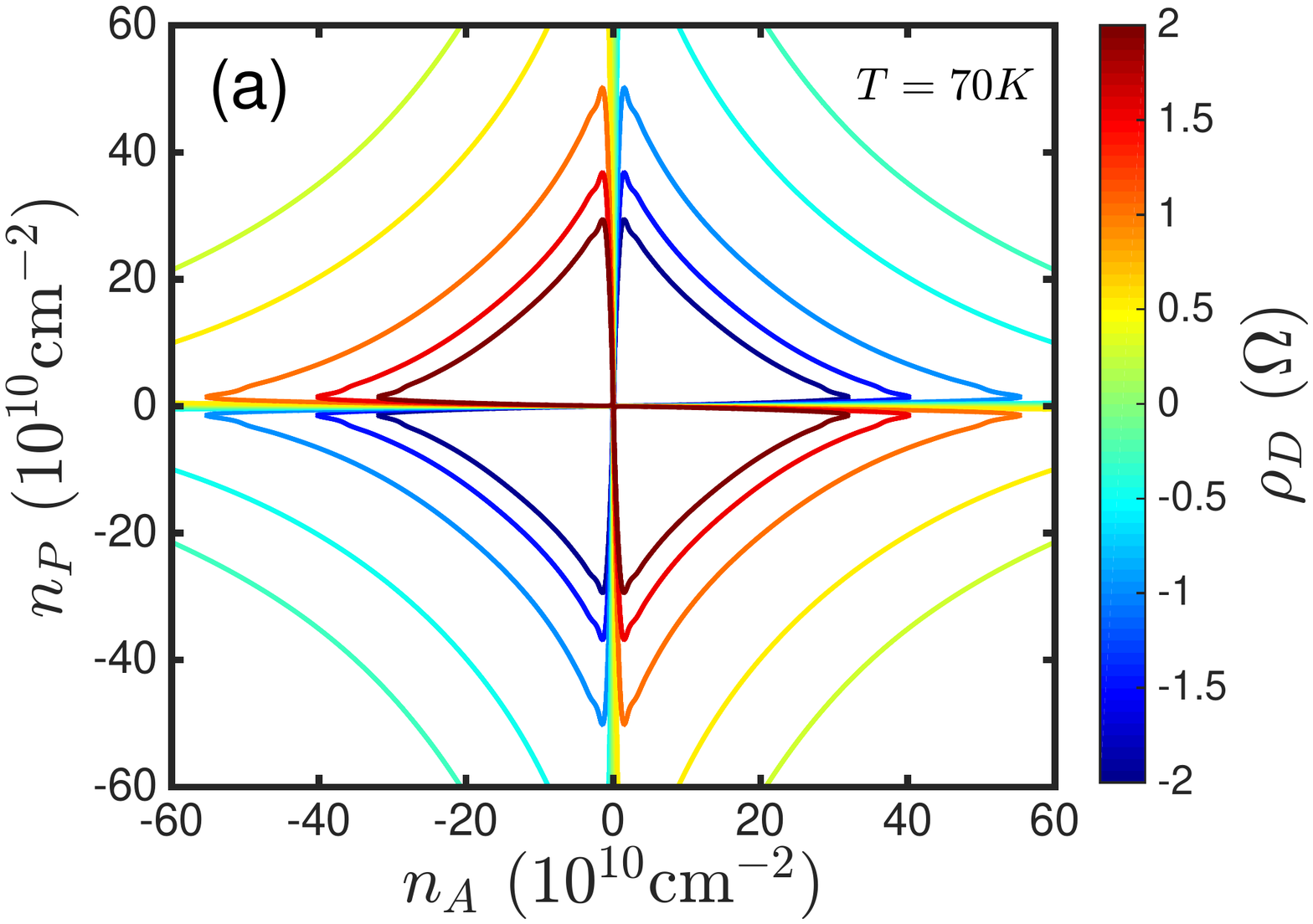}  \\
		\includegraphics[trim=0cm 6.5cm 0cm 6.5cm,clip=true,height=! ,width=8cm] {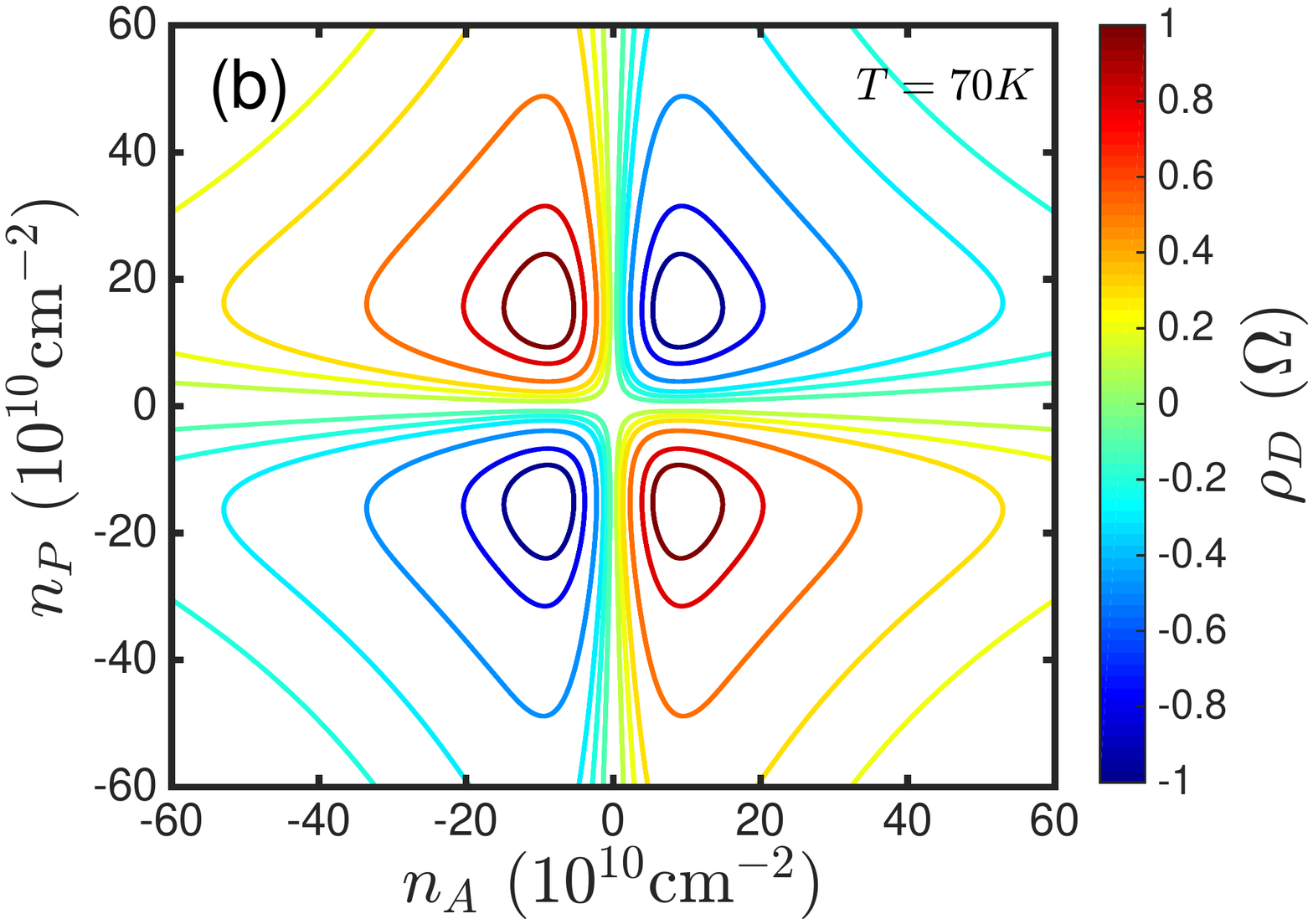}  
		\end{array}$
	\end{center}
	\caption{(color online) Influence of inhomogeneity on drag resistivity contour lines. (a) Contour lines of $\rho_{D}$ for the homogeneous case. The contour lines are strongly concave. (b) Contour lines of $\rho_{D}$ in the presence of charge inhomogeneity, with $n_{\mathrm{rms}}^{(A,P)} = (7,14) \times 10^{10} \mathrm{cm}^{-2}$, $\eta=0$. The contour lines are straight.}
	\label{drag-contours}
\end{figure}

We show in Figs. \ref{drag-contours}(a) and (b) our calculations of $\rho_{\drg}$ as a function of $n_{\act}$ and $n_{\pas}$ using homogeneous theory and EMT respectively. The drag isolevels are strongly concave in the homogeneous case, whereas they are essentially straight in the presence of inhomogeneity. This is due to the fact that EMT basically does a form of averaging of the conductivities in density space. 
At high densities much greater than $n_{\mathrm{rms}}^{(\act,\pas)}$, the contour lines start becoming more concave since the influence of inhomogeneity becomes increasingly unimportant at high densities.
These straight contour lines have also been observed in the experimental works of Ref. \cite{gorbachev_strong_2012} and \cite{kim_coulomb_2012}.

\subsection{4. Drag resistivity with correlated inhomogeneities}

Various proposals have been made concerning the existence of correlations in the density fluctuations of the active and passive layers. Gorbachev {\itshape et al.} argue for the existence of anti-correlated fluctuations in which each hole (electron) puddle lies predominantly above an electron (hole) puddle. This occurs if the density fluctuations arise from strain-induced corrugations in the graphene sheets \cite{gibertini_electron-hole_2012} and the sheets deform themselves so as to minimize the electrostatic potential energy. On the other hand, correlated fluctuations in which each hole (electron) puddle lies predominantly above another hole (electron) puddle have also been suggested \cite{song_energy-driven_2012}. This situation occurs if the fluctuations arise from charged impurities in the surrounding environment \cite{adam_self-consistent_2007} since the puddles in the two layers experience potentials arising from one and the same set of charges. Lastly, the density fluctuations tend toward being completely uncorrelated as the interlayer spacing becomes large.

\begin{figure}[h!]
	\begin{center}$
		\begin{array}{cc}
		\includegraphics[trim=0.8cm 6.5cm 0.8cm 6.5cm,clip=true,height=! ,width=4.4cm] {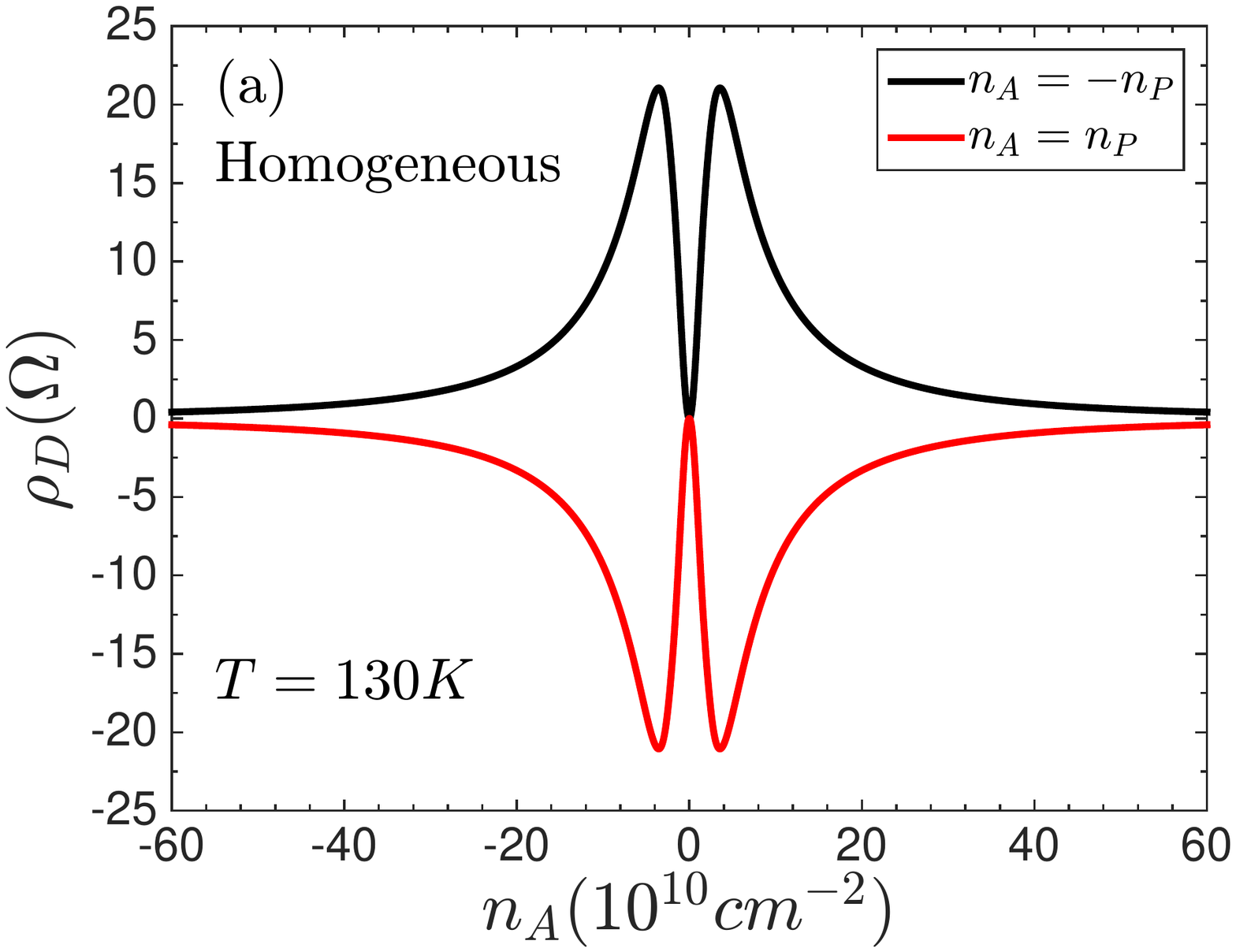}  &
		\includegraphics[trim=0.8cm 6.5cm 0.8cm 6.5cm,clip=true,height=! ,width=4.4cm] {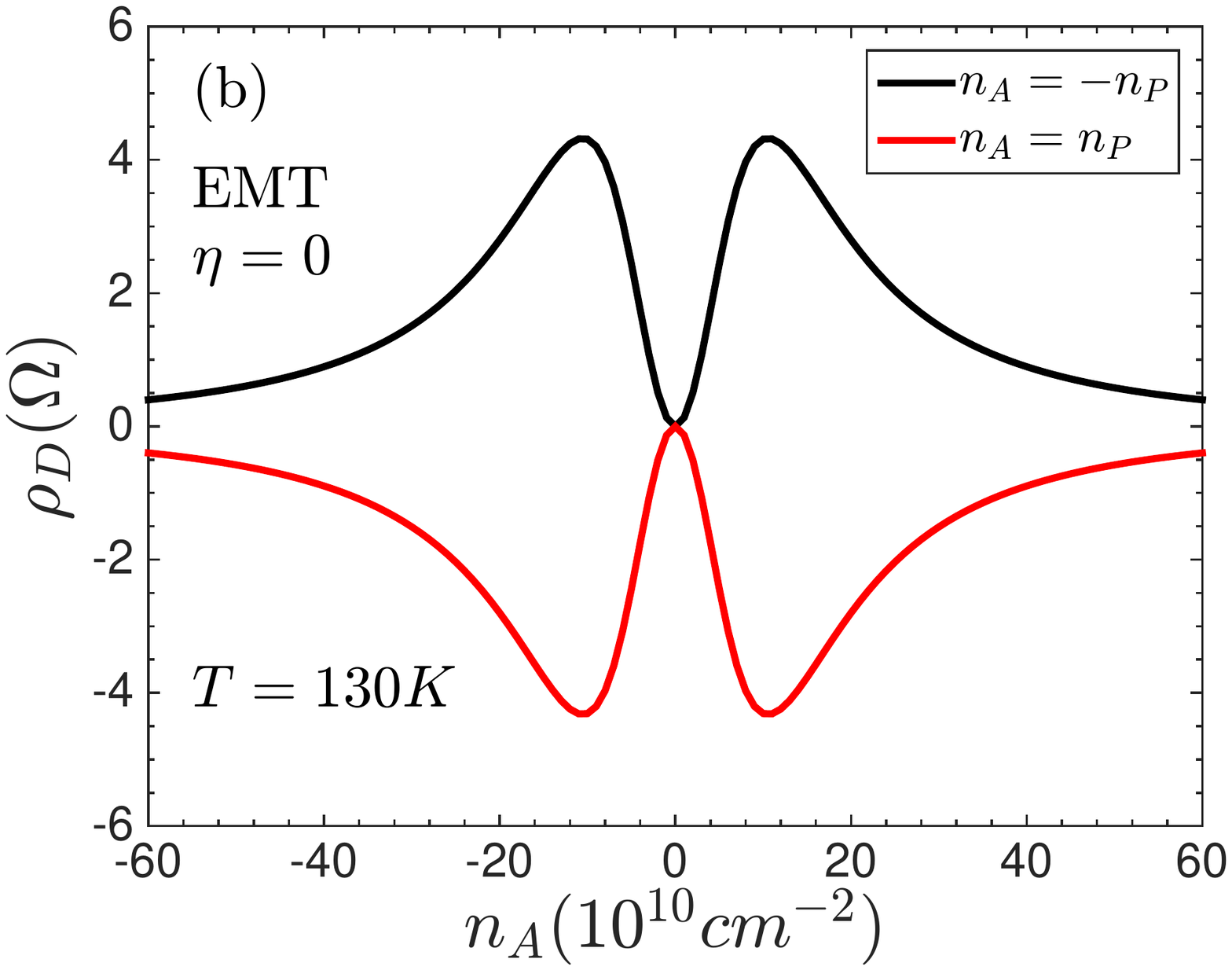}  \\
		\includegraphics[trim=0.8cm 6.5cm 0.8cm 6.5cm,clip=true,height=! ,width=4.4cm] {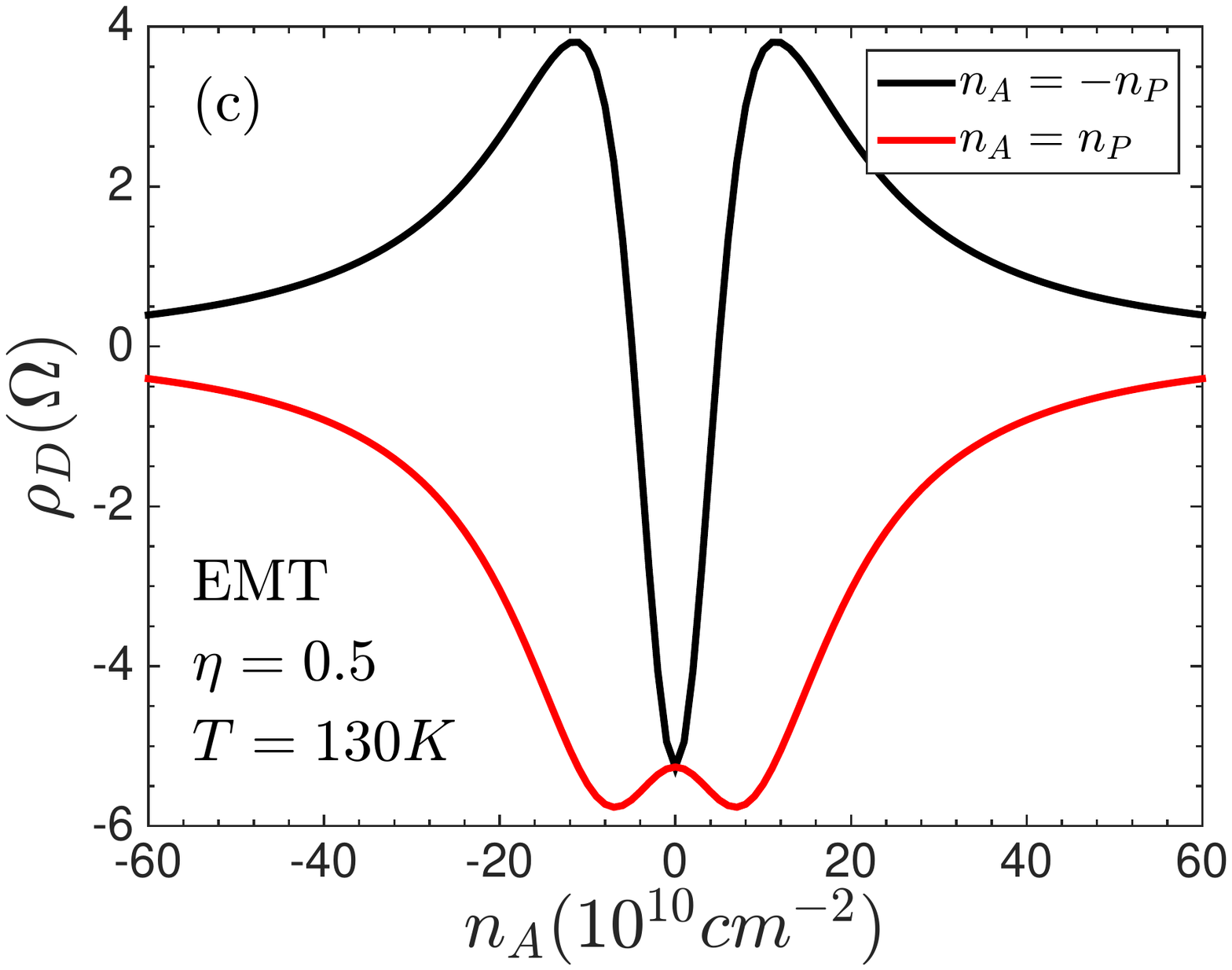}  &
		\includegraphics[trim=0.8cm 6.5cm 0.8cm 6.5cm,clip=true,height=! ,width=4.4cm] {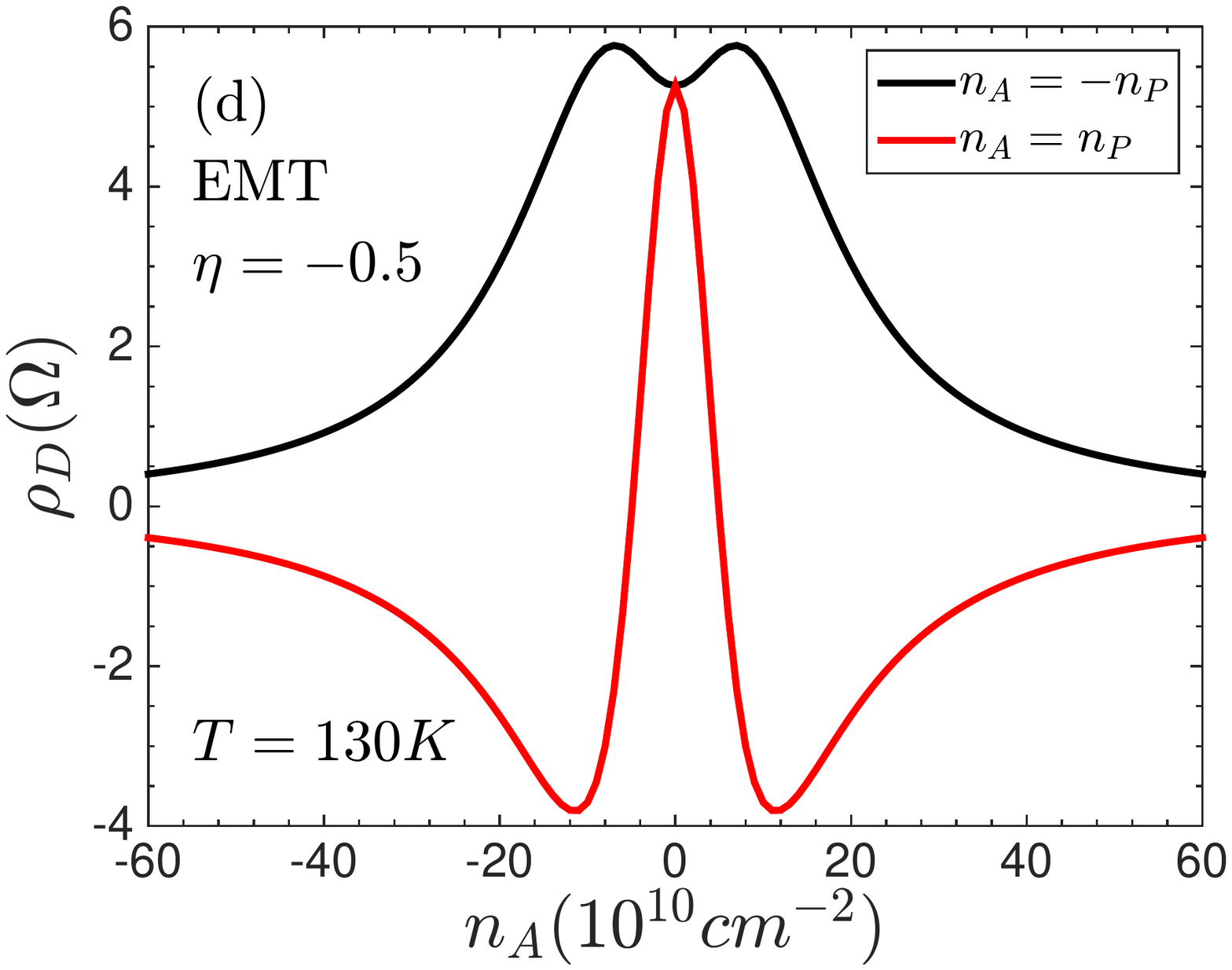}
		\end{array}$
	\end{center}
	\caption{(color online) The role of interlayer correlations on drag resistivity. $\rho_{D}$ calculated as a function of charge density along the lines of oppositely matched densities (upper) and equally matched densities (lower) in the case of (a) perfect charge homogeneity (b) uncorrelated puddles $\eta=0$ (c) correlated puddles $\eta=0.5$ (d) anti-correlated pudles $\eta=-0.5$. Root mean square density fluctuations $n_{\mathrm{rms}}^{(\act,\pas)} = (6,12) \times 10^{10} \mathrm{cm}^{-2}$ throughout.  }
	\label{drag-cuts}
\end{figure}

We study the effect of all three possible types of correlation on $\rho_{\drg}$ and compare with the homogeneous theory. Figs. \ref{drag-cuts}(a) to (d) show $\rho_{\drg}$ for completely homogeneous samples and three different values of correlation coefficient $\eta$ respectively. Strictly speaking, $\eta$ changes as a function of density (and temperature) but we assume it to be constant since we are interested mainly in the qualitative effect of correlations that have more to do with the sign of $\eta$ rather than its exact value. 
With the exception of the double neutrality point (DNP), inhomogeneity always causes a decrease in magnitude of $\rho_{\drg}$, regardless of the nature of correlation. 
This is because drag EMT performs an averaging of the homogeneous $\rho_{\drg}$ in density space, as mentioned at the end of Sec. II. 
This averaging can only lead to an increase in magnitude of $\rho_{\drg}$ at minima of $| \rho_{\drg} |$, and the only one such minima that occurs is located at the double neutrality point \cite{narozhny_coulomb_2012}. 

The presence of non-zero correlations also leads to a finite $\rho_{\drg}$ at the double neutrality point, with correlation (anti-correlation) leading to a negative (positive) drag. This is to be expected since drag between sheets of the same (opposite) sign of charge density is negative (positive). Gorbachev {\itshape et al.} report measuring positive $\rho_{\drg}$ at the double neutrality point in their experiment and that this positive drag always appeared in the regime of $n_{\act},n_{\pas} \sim n_{\mathrm{rms}}$.
There are two proposed explanations for this.
Song and Levitov \cite{song_energy-driven_2012} have demonstrated that energy exchange between the two layers in the presence of correlated fluctuatons yields positive $\rho_{\drg}$ due to thermoelectric effects and suggest that this might explain experiment. They refer to this effect as `energy drag' and to the standard Coulomb force-mediated drag considered in this work (i.e. Eq. (\ref{sigmaD})) and many others \cite{tse_theory_2007,peres_coulomb_2011,narozhny_coulomb_2012,carrega_theory_2012,amorim_coulomb_2012} as `momentum drag'. Song and Levitov ignore the negative contribution of correlated momentum drag at the DNP (cf. Fig. \ref{drag-cuts}(c)).
On the other hand, Gorbachev {\itshape et al.} ignore energy drag and explain their experiment by considering only momentum drag at the DNP and arguing (without the use of drag EMT) that anti-correlated fluctuations arising from random strain in the graphene sheets give rise to positive drag (cf. Fig. \ref{drag-cuts}(d)).
In the presence of nonzero correlations, momentum and energy drag compete at the DNP and a full analysis involving both must be performed in order to predict $\rho_{\drg}$. Such an analysis has not been performed and the exact origin of positive drag at the DNP remains an open problem.
The drag EMT in this work does not include energy drag and is thus unable to make quantitative predictions of drag at the DNP. 
However, it might still be possible to shed some light on the nature of correlations between the two sheets away from the DNP, where $n_{\act},n_{\pas} \gg n_{\mathrm{rms}}$ and energy drag is negligible. 
This is done by measuring $\rho_{\drg}$ along the two lines $n_{\act}=\pm n_{\pas}$ and comparing the magnitude of the drag peaks (i.e. the peaks at finite density away from DNP) along the two lines. 
As shown in Figs. \ref{drag-cuts}(c) and (d), drag EMT predicts that correlated (anti-correlated) fluctuations lead to drag peaks of larger magnitude along the $n_{\act}= n_{\pas}$ ($n_{\act}= -n_{\pas}$) line than the $n_{\act}= -n_{\pas}$ ($n_{\act}= n_{\pas}$) line. Fig. \ref{drag-cuts}(b) shows that uncorrelated fluctuations lead to peaks of equal magnitude along the two lines. 
This suggests a means of deducing the nature of correlations experimentally.
We note that this should be done at low temperatures since inhomogeneity effects become weak at high temperatures.

\section{V. DISCUSSION}


Coulomb drag is a favored probe for studying electron-electron interactions but a standard framework for studying drag in the presence of sample inhomogeneity has been absent till this work.
We have generalized effective medium theory to include Coulomb drag, thus providing for the first time a systematic means of incorporating spatial density fluctuations into drag calculations. 
The importance of the formalism was demonstrated through several examples. 
Standard theory neglecting inhomogeneity suggests that an indirect exciton condensate possesses infinite drag resistivity at zero temperature \cite{vignale_drag_1996}. 
We showed that the presence of density fluctuations yields a finite drag resistivity with a value determined by the amplitude of the fluctuations. 
In the case of drag between graphene sheets, we demonstrated that inhomogeneity is crucial for explaining the experimental observations made in the Manchester experiment of Ref. \cite{gorbachev_strong_2012}. 
We showed that the rise of the drag resistivity peaks with temperature is a direct result of inhomogeneity and cannot be explained by the standard (homogeneous) theory. 
Our calculations also yield isodrag lines that bear a striking resemblance to those from the experiment.
Lastly, there is an ongoing controversy concerning the sign of correlation between the density fluctuations of the two layers. We showed that this may be deduced from drag measurements along different lines in the $(n_{\act},n_{\pas})$ density space. 

Several future avenues of research arise from this work. 
First, we have considered only the Coulomb-mediated momentum exchange between the layers and ignored interesting proposals of possible thermoelectric effects \cite{lung_thermoelectric_2011,song_energy-driven_2012}. 
In principle, it is possible to formulate a generalized effective medium theory of drag incorporating thermoelectricity along the lines of Ref. \cite{webman_thermoelectric_1977}. 
Second, it would be of great interest to explore the effects of inhomogeneity in double layer graphene in the hydrodynamic regime and also in different drag setups such as bilayer graphene \cite{li_negative_2016,lee_giant_2016}, topological insulators \cite{efimkin_drag_2013}, two-dimensional electron gases \cite{de_cumis_complete_2016} and hybrid graphene-two-dimensional gas setups \cite{gamucci_anomalous_2014}. 
Third, the present work may be generalized to obtain a magnetodrag EMT (i.e. an EMT for Coulomb drag in the presence of a magnetic field).
Such a theory is at present still lacking in the literature and would be especially important because to date, successful observations of exciton condensation have typically involved non-zero magnetic fields \cite{nandi_exciton_2012,li_excitonic_2016,liu_quantum_2016}. 
Last, it would be interesting to implement the dielectric inhomogeneity \cite{badalyan_enhancement_2012} and virtual phonon \cite{amorim_coulomb_2012-1} enhancement mechanisms in theoretical calculations to obtain precise agreement with experiment.
We have also assumed for the sake of simplicity that $\eta$ and $n_{\mathrm{rms}}^{(A,P)}$ are effectively constant as functions of density and temperature.
This assumption does not affect the qualitative accuracy of the results but nonetheless has a quantitative effect.
One possible way to remedy this is to use the rigorous theory presented in Ref. \cite{rodriguez-vega_ground_2014} for calculating $\eta$ and $n_{\mathrm{rms}}^{(A,P)}$ as a function of the system parameters.
We leave these as problems for future work.

\section{Acknowledgment}
We are grateful for very useful discussions with Cory Dean, Eugene Mele, Boris Narozhny, Justin Song, Marco Polini, Navneeth Ramakrishnan, and Giovanni Vignale. D.H.~also thanks Lim Yu Chen for his work as part of a high school project in developing early versions of the Matlab codes used here. 
This work was supported by the National Research Foundation Singapore under its fellowship program (NRF-NRFF2012-01) and by the Singapore Ministry of Education and Yale-NUS College through Grant No.~R-607-265-01312. 
BYKH acknowledges the Professional Development Leave granted by the University of Akron and the hospitality of the Center for Advanced 2D Materials at the National University of Singapore.
The authors also gratefully acknowledge the use of the dedicated computational facilities at the Centre for Advanced 2D Materials and the invaluable assistance of Miguel Dias Costa in making use of these resources.



\renewcommand{\theequation}{A\arabic{equation}}


\setcounter{equation}{0}

\section{APPENDIX A: DERIVATION OF SINGLE LAYER EMT}
\begin{figure}
	\begin{center}$
		\begin{array}{c}
		\includegraphics[trim=0cm 3cm 0cm 3cm,clip=true,height=! ,width=9cm] {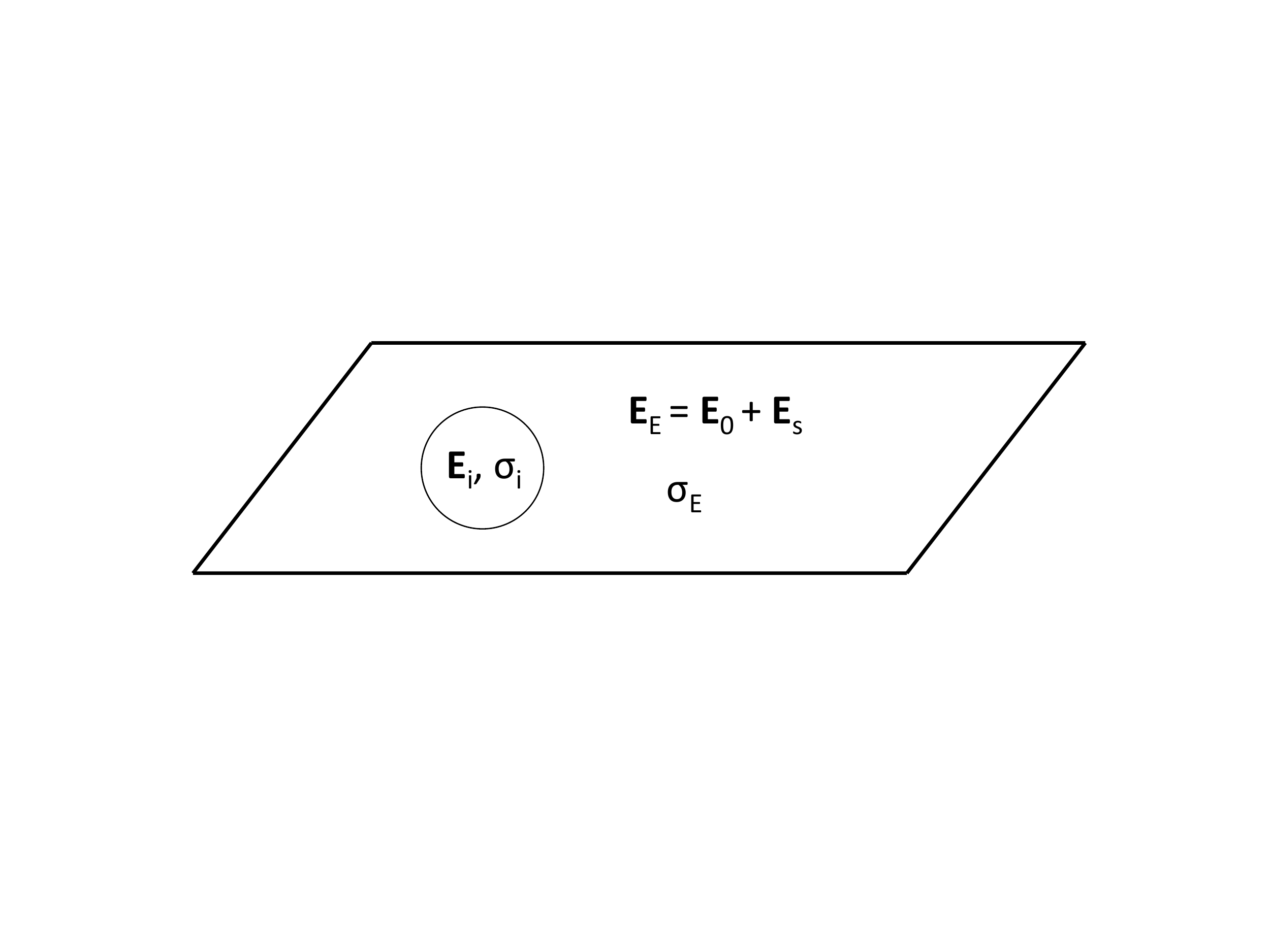}  
		\end{array}$
	\end{center}
	\caption{A circular region of inhomogeneity embedded inside a homogeneous effective medium of conductivity $\sigma_\eff$.}
	\label{single-layer-diagram}
\end{figure}

For completeness, we also review the derivation of the monolayer EMT result, Eq.~(\ref{sigmaEMT-mono}). 
Useful reviews of monolayer EMT may be found in 
Consider a sheet of 2D material which is made up of a patchwork of $N$ areas (i.e.~puddles) each with differing conductivities $\sigma_{i}$, where $i =1 , \cdots , N$. The areal fraction of the $i$th puddle is denoted by $f_{i}$ with $\sum_{i} f_{i} = 1$. We wish to calculate the effective conductivity of this sheet. We first imagine that each puddle is embedded in a homogeneous effective medium of conductivity $\sigma_\eff$ and through which permeates a uniform electric field $\vec{E}_{0}$. Next, we work out the electric field $\vec{E}_{i}$ inside each puddle in the form $\vec{E}_{i} = ( \cdots ) \vec{E}_{0}$. 
Consider the $i$th puddle embedded in the effective medium as shown in Fig.~\ref{single-layer-diagram}. For simplicity, we assume that all puddles are circles of radius $a$. We find the electric field inside this puddle. 

The conductivity of the medium is $\sigma_\eff$ and it is permeated by a uniform field $\vec{E}_{0} = E_{0} \vec{e}_x$ corresponding to an external field that we will refer to as the primary field. We denote the field inside the puddle as $\vec{E}_{i}$ and the field outside as $\vec{E}_\eff$. We assume that the puddle possesses a uniform polarization $\vec{M}$ pointing in the same direction as the external field, so that $\vec{M} = M \vec{e}_x$. From electrostatics, such a circular puddle is associated with an electric field in the region outside it given by
\begin{equation}
E_{s}(\vec{r}) = \frac{a^{2}}{2 \epsilon_{0} r^{2}} \left[ 2 ( \vec{M} \cdot \vec{e}_{r} ) \vec{e}_{r} - \vec{M} \right] \label{Es-field}
\end{equation}
with a corresponding potential 
\begin{equation}
U_{s}(\vec{r}) = \frac{a^{2}}{2 \epsilon_{0}} \frac{M}{r} \cos(\theta).
\end{equation}
Here, the subscript `s' is short for `secondary' and we placed the origin of our radial coordinate system at the center of the circular puddle. Hence, the total field outside ($r > a$) is, taking the $\vec{e}_{r}$ components, 
\begin{eqnarray}
E_{E,r} (r,\theta) &=&  E_{0,r} + E_{s,r} \\ \nonumber 
&=& E_{0} \cos(\theta) + \frac{a^{2}}{2 \epsilon_{0} r^{2}} M \cos(\theta)
\end{eqnarray}
and the potential is 
\begin{equation}
U_\eff (r,\theta) = - E_{0} r \cos(\theta) +\frac{a^{2}}{2\epsilon_{0}}\frac{M}{r} \cos(\theta).
\end{equation}
Next, we consider the field inside the puddle. We guess that the field inside is proportional to the externally applied field $\vec{E}_{0}$. Hence,
\begin{eqnarray}
\vec{E}_{i} &=& C \vec{E}_{0} \\ \nonumber
&=& C E_{0} \vec{e}_x \\ \nonumber
&=& C E_{0} \cos(\theta) \vec{e}_{r} \label{guess}
\end{eqnarray}
with potential 
\begin{equation}
U_{i} (r,\theta) = - C E_{0} r \cos(\theta). \label{guess-pot}
\end{equation}
We now use boundary conditions to solve for the unknown $C$, and obtain the desired field inside the puddle. The boundary conditions that must be obeyed at the boundary of the puddle are the continuity of potential, and continuity of radial current density. In equations, they are
\begin{equation}
U_\eff (a,\theta) = U_{i} (a,\theta)
\end{equation}
and 
\begin{equation}
\sigma_\eff E_{E,r} = \sigma_{i} E_{i,r}.
\end{equation}
We can make use of these two boundary conditions to solve for the two unknowns $M$ and $C$. This yields 
\begin{equation}
C = \frac{2\sigma_\eff}{\sigma_\eff + \sigma_{i}}
\end{equation}
and 
\begin{equation}
M = 2 \epsilon_{0} E_{0} \left(\frac{2\sigma_\eff}{\sigma_\eff + \sigma_{i}} -1 \right).
\end{equation}
The fact that solutions for $C$ and $M$ exist validates our earlier guess in Eq.~(\ref{guess}), which we know is the unique physical solution due to the uniqueness theorem. We are not interested in $M$ here. Our objective was to find $\vec{E}_{i}$, which we have successfully done by determining $C$. Explicitly, we have found that 
\begin{equation}
{E}_{i} = \left( \frac{2\sigma_\eff}{\sigma_\eff + \sigma_{i}} \right) E_{0}, \label{E-in}
\end{equation}
where $E_{i}$ is the magnitude of $\vec{E}_{i}$. 
We substitute the above into the EMT self-consistency condition
\begin{equation}
\sum_{i=1}^{N} f_{i} \vec{E}_{i} = \vec{E}_{0}, \label{EMT-self-con}
\end{equation}
and simplify to obtain
\begin{equation}
\sum_{i=1}^{N} f_{i} \cdot \frac{\sigma_{i} - \sigma_\eff}{\sigma_{i} + \sigma_\eff } = 0.
\end{equation}
Generalizing to a continuum of puddles and denoting the continuous puddle label by $n$, one obtains Eq.~(\ref{sigmaEMT-mono}).

\renewcommand{\theequation}{B\arabic{equation}}
\setcounter{equation}{0}

\section{APPENDIX B: PROOF OF ONSAGER RECIPROCITY}

In the context of drag, Onsager reciprocity \cite{casimir_onsagers_1945} predicts that there should be no difference in drag resistivity when the active and passive layers are switched.
We prove that Onsager reciprocity is obeyed by the drag EMT derived in this work.
Mathematically, proving Onsager reciprocity amounts to interchanging all `A' and `P' superscripts (denoted henceforth by `$\textsc{A} \leftrightarrow \textsc{P}$') and showing that drag resistivity remains the same.
Since it is clear that interchanging $\sigma_{\act}^{\eff}$ and $\sigma_{\pas}^{\eff}$ in the drag resistivity expression Eq. (\ref{rhoD}) makes no difference, our remaining task is to prove smilar invariance for $\sigma_{\drg}^{\eff}$. 
We start by considering the discretized version of $\sigma_{\drg}^{\eff}$ in Eq. (\ref{sigmaEMT-drag}), which may be rewritten as 
\begin{equation*}
\sigma_{\drg}^{\eff} = \frac{\mathrm{Numerator}}{\mathrm{Denominator}}, 
\end{equation*}
where 
\begin{equation}
\mathrm{Numerator} = \sum_{i} f_{i} \cdot \frac{ \sigma^\drg_{i} }
{(\sigma^\act_\eff+\sigma^\act_{i})(\sigma^\pas_\eff+\sigma^\pas_{i})}, \label{numerator}
\end{equation}
and 
\begin{equation}
\mathrm{Denominator}=\sum_{i} f_{i} \cdot \frac{ \frac{\sigma^\act_{i}}{\sigma^\act_\eff} }
{(\sigma^\act_\eff+\sigma^\act_{i})(\sigma^\pas_\eff+\sigma^\pas_{i})}. \label{denominator}
\end{equation}
The numerator Eq. (\ref{numerator}) is invariant under $\textsc{A} \leftrightarrow \textsc{P}$ because of the invariance of the standard (homogeneous) drag conductivity, as seen from Eq. (\ref{sigmaD}).
To prove that the denominator is also invariant, we take the difference of the monolayer EMT equations  (\ref{sigmaEMT-active}) and (\ref{sigmaEMT-passive}),
\begin{equation*}
\sum_{i} f_{i} \cdot \frac{(\sigma^\act_{i}-\sigma^\act_\eff)(\sigma^\pas_\eff+\sigma^\pas_{i})- (\sigma^\pas_{i}-\sigma^\pas_\eff)(\sigma^\act_\eff+\sigma^\act_{i}) } 
{(\sigma^\act_\eff+\sigma^\act_{i})(\sigma^\pas_\eff+\sigma^\pas_{i})}  =0. 
\end{equation*}
Simplifying and rearranging, we obtain 
\begin{equation}
\sum_{i} f_{i} \cdot \frac{ \frac{\sigma^\act_{i}}{\sigma^\act_\eff} }
{(\sigma^\act_\eff+\sigma^\act_{i})(\sigma^\pas_\eff+\sigma^\pas_{i})} = \sum_{i} f_{i} \cdot \frac{ \frac{\sigma^\pas_{i}}{\sigma^\pas_\eff} }
{(\sigma^\act_\eff+\sigma^\act_{i})(\sigma^\pas_\eff+\sigma^\pas_{i})}. 
\end{equation}
This proves that Eq. (\ref{denominator}) is also invariant under $\textsc{A} \leftrightarrow \textsc{P}$. 
We have thus established that the discrete version of $\sigma_{\drg}^{\eff}$ obeys Onsager reciprocity. 
Taking the continuum limit, we conclude that the continuous version Eq. (\ref{sigmaEMT-drag-cont}) does too and $\sigma_{\drg}^{\eff}$ remains unchanged when the roles of the two layers are swapped. 
This completes the proof of Onsager reciprocity.

\renewcommand{\theequation}{C\arabic{equation}}
\setcounter{equation}{0}

\section{APPENDIX C: HOMOGENEOUS DRAG THEORY}

\subsection{1. Drag Conductivity Expressions}

The dynamically screened interlayer Coulomb interaction is given by
\begin{equation}
V(q,\omega,d) = \frac{V_{12}(q,d)}{\epsilon_{\drg} (q,\omega ,d) }
\end{equation}
where the double-layer dielectric function $\epsilon_{\drg}$ is given by
\begin{widetext}
\begin{equation}
\epsilon_{\drg} (q,\omega ,d,T) = \left( 1 - V_{11}(q) \Pi_{\act} (q,\omega,T) \right) \left( 1 - V_{22}(q) \Pi_{\pas}(q,\omega,T) \right) - V_{12}(q)V_{21}(q)\Pi_{\act}(q, \omega, T) \Pi_{\pas} (q, \omega, T), \label{dielec-dbl}
\end{equation}
\end{widetext}
with bare interlayer and intralayer Coulomb potentials $V_{12}(q,d) = V_{21}(q,d) = 2\pi e^{2}\exp(-qd) / \kappa q$ and $V_{11}(q)=V_{22}(q) = 2\pi e^{2} / \kappa q$ respectively, where $\kappa$ is the dielectric constant of the material encapsulating the graphene sheets and $d$ is the interlayer spacing. We assume $d=9 \mathrm{nm}$ in all our calculations here, corresponding to the sample measured in Fig 3(a) of Gorbachev {\itshape et al}. $\Pi_{i}$ is the dynamical polarizability of layer $i$ as given in Ref.~\cite{ramezanali_finite-temperature_2009}. 

The physical meaning of the nonlinear susceptibility is that it is a response function relating the voltage felt across each layer with the current it induces within it \cite{narozhny_coulomb_2012} via
\begin{equation}
\mathbf{i}(\omega) = \int d\mathbf{r}_{1} \int d\mathbf{r}_{2} \mathbf{\Gamma}(\omega,\mathbf{r}_{1},\mathbf{r}_{2}) V(\mathbf{r}_{1}) V(\mathbf{r}_{2}).
\end{equation}
The $x$-component of the nonlinear susceptibility of layer $i$ is given by 
\begin{eqnarray}
\Gamma ^x _{i} \left(n_{i}, \mathbf{q} ,\omega \right)  &\equiv&  \Gamma ^x_{i}(\omega, \mathbf{q}, \mu_{i} /k_{\scr{B}}T) \\ \nonumber 
&=&  \Gamma_{i}(\omega, q, \mu_{i} /k_{\scr{B}}T) \cos (\theta_{q}),
\end{eqnarray} 
where we convert charge density to chemical potential using the methods detailed in the next subsection. It is convenient at this stage to switch to dimensionless notation,
\begin{equation}
\tilde{\omega} = \frac{\hbar \omega}{k_{\scr{B}} T}, \hspace{2mm} \tilde{q} = \frac{\hbar v_{F} q}{k_{\scr{B}} T}, \hspace{2mm} \tilde{\mu} = \frac{\mu}{k_{\scr{B}} T}, \hspace{2mm} \tilde{E} = \frac{E}{k_{\scr{B}} T}, \hspace{2mm} z=\frac{2\tilde{E}+\tilde{\omega}}{\tilde{q}}.
\end{equation}
With this notation, we follow the approach detailed in Ref.~\cite{narozhny_coulomb_2012} to obtain the expressions 
\begin{widetext}
\begin{eqnarray}
\Gamma_{i}(\omega, q, \frac{\mu_{i}}{k_{\scr{B}}T}) &=& -\frac{4e}{\hbar v_{F}} \tilde{\Gamma}_{i} ( \tilde{\omega}, \tilde{q}, \tilde{\mu_{i}}), \nonumber \\
\tilde{\Gamma}_{i} ( \tilde{\omega}, \tilde{q}, \tilde{\mu}_{i}) &=& \frac{1}{4\pi} G(\tilde{\omega},\tilde{q},\tilde{\mu_{i}}) \tilde{q}, \nonumber \\
G(\tilde{\omega},\tilde{q},\tilde{\mu_{i}}) &=& 
\begin{cases}
-\frac{1}{2} \int ^{1}_{0} dz I(z,\tilde{q},\tilde{\omega},\tilde{\mu_{i}}) \sqrt{\frac{1-z^{2}}{\frac{\tilde{\omega}^{2}}{\tilde{q}^{2}}-1 }} K_{i}(z,\tilde{q},\tilde{\omega}), & | \tilde{\omega}| > \tilde{q},  \\ 
\frac{1}{2} \int ^{\infty}_{1} dz I(z,\tilde{q},\tilde{\omega},\tilde{\mu}_{i}) \sqrt{\frac{z^{2}-1}{ 1-\frac{\tilde{\omega}^{2}}{\tilde{q}^{2}} }} K_{i}(z,\tilde{q},\tilde{\omega}), & | \tilde{\omega} | < \tilde{q}, 
\end{cases} \nonumber \\
I(z,\tilde{q},\tilde{\omega},\tilde{\mu}_{i})  &=& \tanh\left(\frac{z\tilde{q}-\tilde{\omega}-2\tilde{\mu}_{i}}{4}\right) - \tanh\left(\frac{z\tilde{q}+\tilde{\omega}-2\tilde{\mu}_{i}}{4}\right) + \tanh \left(\frac{z\tilde{q}+\tilde{\omega}+2\tilde{\mu}_{i}}{4}\right)-\tanh\left(\frac{z\tilde{q}-\tilde{\omega}+2\tilde{\mu}_{i}}{4}\right), \nonumber \\
K_{i}(z,\tilde{q},\tilde{\omega}) &=& \tilde{\tau}_{i} \left(\frac{z\tilde{q}-\tilde{\omega}}{2}\right) \frac{z\tilde{\omega}-\tilde{q}}{z\tilde{q}-\tilde{\omega}} - \tilde{\tau}_{i}\left(\frac{-z\tilde{q}-\tilde{\omega}}{2}\right)\frac{z\tilde{\omega}+\tilde{q}}{z\tilde{q}+\tilde{\omega}}, \nonumber \\
\tilde{\tau}_{i}(\tilde{E}) &=& \frac{k_{\scr{B}} T \tau_{i}(E)}{\hbar}. \label{NLS}
\end{eqnarray}
\end{widetext}
where $\tau_{i}$ is the transport scattering time of layer $i$. Note however that the dimensionless frequency and momenta defined in this work differ from that in Ref. \cite{narozhny_coulomb_2012} by a factor of $1/2$. In this paper we assume that electron-charged impurity scattering dominates over all other scattering mechanisms. In this case, $\tau_{i}$ is given by
\begin{eqnarray}
\frac{1}{\tau_{i}(E)} &=& \frac{4\pi n_\mathrm{imp}^{(i)}}{\hbar} \int \frac{d^{2}k'}{(2\pi)^{2}} \left| \frac{V_\mathrm{imp}(q,d_{\mathrm{imp}}^{(i)})}{\epsilon_{s}(q)} \right|^{2}  \times \\ \nonumber
&&  \frac{1-\cos^{2}(\theta_{\mathbf{k,k'}})}{4} \delta (E_{k} -E_{k'}), \label{tau-imp}
\end{eqnarray}
where $q = |\mathbf{k} - \mathbf{k}'|$ and $\theta_{\mathbf{k,k'}}$ is the angle between the initial and final wave vectors $\mathbf{k}$ and $\mathbf{k'}$ in a scattering event. $n_\mathrm{imp}^{(i)}$ is the areal concentration of charged impurities in an impurity plane located at a distance $d_{\mathrm{imp}}^{(i)}$ away from the graphene sheet. $V_\mathrm{imp} (q) = 2 \pi e^{2} / (\kappa q) \exp(- q d_\mathrm{imp}^{(i)}) $ where $d_\mathrm{imp}^{(i)}$ is the distance between the graphene sheet and the charged impurities which are assumed to lie in a single plane. In our calculations, we assume that both layers see one and the same impurity plane so that $n^\act_{\mathrm{imp}} = n^\pas_{\mathrm{imp}} \equiv n_\mathrm{imp}$. In the experiment of Gorbachev {\itshape et al.}, the impurity plane is expected to be near the $\mathrm{SiO_{2}}$ wafer on which the drag heterostructure rests. Hence, we assume $d_{\mathrm{imp}}^{\pas}=10 \mathrm{nm}$ and $d_{\mathrm{imp}}^{\act}=20 \mathrm{nm}$ corresponding to the fact that the passive layer lies nearer the $\mathrm{SiO_{2}}$ wafer in Gorbachev {\itshape et al.}

The single layer dielectric function $\epsilon_{s}$ is given by 
\begin{equation}
\epsilon_{s}(q) = 1 - \Pi(q,T) V_{11}(q),  \label{dielec-sgl}
\end{equation}
where $\Pi(q,T)$ is the static polarizability of graphene. As pointed out in Ref.~\cite{narozhny_coulomb_2012}, the nonlinear susceptibility contains a logarithmic divergence along the line $\tilde{q} = \tilde{\omega}$ so long as $\tau$ has an $E$-dependence. In this work, we prevent $\sigma_{\drg}$ from diverging in calculations by using the dynamical polarizability of Ref.~\cite{ramezanali_finite-temperature_2009} in the dielectric function. This introduces a divergence in the denominator of $\sigma_{\drg}$ which cancels that in the numerator, leaving behind a finite and well-defined quantity. We shall assume the value of the graphene `fine structure constant' is $r_\mathrm{s} = 0.568 $, corresponding to an estimated value of $\kappa =3.5$ for hexagonal boron nitride. The above expressions constitute all the ingredients one needs to calculate the drag homogeneous conductivity. 
	
\subsection{2. Relation between Charge Density and Chemical Potential}

Here we review the one-to-one correspondence between charge density and chemical potential given a fixed temperature. The charge density is defined as $n= n_{e} - n_{h}$ where $n_{e}$ and $n_{h}$ refer to the electron and hole densities respectively. These densities are obtained from the chemical potential $\mu$ and temperature $T$ via
\begin{equation}
n_{e,h}=-\frac{2}{\pi}\left(\frac{k_{B}T}{\hbar v_{F}}\right)^{2}\mathrm{Li}_{2}\left[-\exp\left(\pm\frac{\mu}{k_{B}T}\right)\right], \label{neh}
\end{equation}
where $\mathrm{Li}_{2}$ refers to the dilogarithm function. This equation allows us to find the charge density given the chemical potential and temperature. 

We can also find the chemical potential given the charge density and temperature. This is done by noting that $E_{F} =\hbar v_{F} \sqrt{\pi |n|} \hspace{0.5mm} \mathrm{sign}(n)$ and using the relation 
\begin{equation}
\frac{\mu}{E_{F}} = F_{\mu} \left(  \frac{k_{\scr{B}} T}{|E_{F}|}  \right), \label{Fmu}
\end{equation}
where
\begin{equation}
F_{\mu} (x) = \bar{g} (x) \left( 1 - \frac{\pi^{2} x^{2}}{6} \right) + g(x) / \left[ 4\log(2) x \right],
\end{equation}
with $g(x) = (1+ \mathrm{Erf}[10(x-0.5)] )/2 $ and $\bar{g} (x) = \mathrm{Erfc}[10(x-0.5)]/2 $. Eqs (\ref{neh}) and (\ref{Fmu}) allow one to trivially convert a function of charge density to a function of chemical potential and vice versa. 	
	
\bibliographystyle{apsrev4-1.bst}
\bibliography{MyLibrary.bib}

\begin{thebibliography}{48}%
\makeatletter
\providecommand \@ifxundefined [1]{%
 \@ifx{#1\undefined}
}%
\providecommand \@ifnum [1]{%
 \ifnum #1\expandafter \@firstoftwo
 \else \expandafter \@secondoftwo
 \fi
}%
\providecommand \@ifx [1]{%
 \ifx #1\expandafter \@firstoftwo
 \else \expandafter \@secondoftwo
 \fi
}%
\providecommand \natexlab [1]{#1}%
\providecommand \enquote  [1]{``#1''}%
\providecommand \bibnamefont  [1]{#1}%
\providecommand \bibfnamefont [1]{#1}%
\providecommand \citenamefont [1]{#1}%
\providecommand \href@noop [0]{\@secondoftwo}%
\providecommand \href [0]{\begingroup \@sanitize@url \@href}%
\providecommand \@href[1]{\@@startlink{#1}\@@href}%
\providecommand \@@href[1]{\endgroup#1\@@endlink}%
\providecommand \@sanitize@url [0]{\catcode `\\12\catcode `\$12\catcode
  `\&12\catcode `\#12\catcode `\^12\catcode `\_12\catcode `\%12\relax}%
\providecommand \@@startlink[1]{}%
\providecommand \@@endlink[0]{}%
\providecommand \url  [0]{\begingroup\@sanitize@url \@url }%
\providecommand \@url [1]{\endgroup\@href {#1}{\urlprefix }}%
\providecommand \urlprefix  [0]{URL }%
\providecommand \Eprint [0]{\href }%
\providecommand \doibase [0]{http://dx.doi.org/}%
\providecommand \selectlanguage [0]{\@gobble}%
\providecommand \bibinfo  [0]{\@secondoftwo}%
\providecommand \bibfield  [0]{\@secondoftwo}%
\providecommand \translation [1]{[#1]}%
\providecommand \BibitemOpen [0]{}%
\providecommand \bibitemStop [0]{}%
\providecommand \bibitemNoStop [0]{.\EOS\space}%
\providecommand \EOS [0]{\spacefactor3000\relax}%
\providecommand \BibitemShut  [1]{\csname bibitem#1\endcsname}%
\let\auto@bib@innerbib\@empty
\bibitem [{\citenamefont {Narozhny}\ and\ \citenamefont
  {Levchenko}(2016)}]{narozhny_coulomb_2016}%
  \BibitemOpen
  \bibfield  {author} {\bibinfo {author} {\bibfnamefont {B.}~\bibnamefont
  {Narozhny}}\ and\ \bibinfo {author} {\bibfnamefont {A.}~\bibnamefont
  {Levchenko}},\ }\bibfield  {title} {\emph {\bibinfo {title} {Coulomb drag},\
  }}\href {\doibase 10.1103/RevModPhys.88.025003} {\bibfield  {journal}
  {\bibinfo  {journal} {Reviews of Modern Physics}\ }\textbf {\bibinfo {volume}
  {88}},\ \bibinfo {pages} {025003} (\bibinfo {year} {2016})}\BibitemShut
  {NoStop}%
\bibitem [{\citenamefont {Hu}\ and\ \citenamefont
  {Flensberg}(1996)}]{hu_electron-electron_1996}%
  \BibitemOpen
  \bibfield  {author} {\bibinfo {author} {\bibfnamefont {B.~Y.-K.}\
  \bibnamefont {Hu}}\ and\ \bibinfo {author} {\bibfnamefont {K.}~\bibnamefont
  {Flensberg}},\ }\bibfield  {title} {\emph {\bibinfo {title}
  {Electron-electron scattering in linear transport in two-dimensional
  systems},\ }}\href {\doibase 10.1103/PhysRevB.53.10072} {\bibfield  {journal}
  {\bibinfo  {journal} {Physical Review B}\ }\textbf {\bibinfo {volume} {53}},\
  \bibinfo {pages} {10072} (\bibinfo {year} {1996})}\BibitemShut {NoStop}%
\bibitem [{\citenamefont {Solomon}\ \emph {et~al.}(1989)\citenamefont
  {Solomon}, \citenamefont {Price}, \citenamefont {Frank},\ and\ \citenamefont
  {La~Tulipe}}]{solomon_new_1989}%
  \BibitemOpen
  \bibfield  {author} {\bibinfo {author} {\bibfnamefont {P.~M.}\ \bibnamefont
  {Solomon}}, \bibinfo {author} {\bibfnamefont {P.~J.}\ \bibnamefont {Price}},
  \bibinfo {author} {\bibfnamefont {D.~J.}\ \bibnamefont {Frank}}, \ and\
  \bibinfo {author} {\bibfnamefont {D.~C.}\ \bibnamefont {La~Tulipe}},\
  }\bibfield  {title} {\emph {\bibinfo {title} {New phenomena in coupled
  transport between 2d and 3d electron-gas layers},\ }}\href {\doibase
  10.1103/PhysRevLett.63.2508} {\bibfield  {journal} {\bibinfo  {journal}
  {Physical Review Letters}\ }\textbf {\bibinfo {volume} {63}},\ \bibinfo
  {pages} {2508} (\bibinfo {year} {1989})}\BibitemShut {NoStop}%
\bibitem [{\citenamefont {Gramila}\ \emph {et~al.}(1991)\citenamefont
  {Gramila}, \citenamefont {Eisenstein}, \citenamefont {MacDonald},
  \citenamefont {Pfeiffer},\ and\ \citenamefont {West}}]{gramila_mutual_1991}%
  \BibitemOpen
  \bibfield  {author} {\bibinfo {author} {\bibfnamefont {T.~J.}\ \bibnamefont
  {Gramila}}, \bibinfo {author} {\bibfnamefont {J.~P.}\ \bibnamefont
  {Eisenstein}}, \bibinfo {author} {\bibfnamefont {A.~H.}\ \bibnamefont
  {MacDonald}}, \bibinfo {author} {\bibfnamefont {L.~N.}\ \bibnamefont
  {Pfeiffer}}, \ and\ \bibinfo {author} {\bibfnamefont {K.~W.}\ \bibnamefont
  {West}},\ }\bibfield  {title} {\emph {\bibinfo {title} {Mutual friction
  between parallel two-dimensional electron systems},\ }}\href {\doibase
  10.1103/PhysRevLett.66.1216} {\bibfield  {journal} {\bibinfo  {journal}
  {Physical Review Letters}\ }\textbf {\bibinfo {volume} {66}},\ \bibinfo
  {pages} {1216} (\bibinfo {year} {1991})}\BibitemShut {NoStop}%
\bibitem [{\citenamefont {Sivan}\ \emph {et~al.}(1992)\citenamefont {Sivan},
  \citenamefont {Solomon},\ and\ \citenamefont
  {Shtrikman}}]{sivan_coupled_1992}%
  \BibitemOpen
  \bibfield  {author} {\bibinfo {author} {\bibfnamefont {U.}~\bibnamefont
  {Sivan}}, \bibinfo {author} {\bibfnamefont {P.~M.}\ \bibnamefont {Solomon}},
  \ and\ \bibinfo {author} {\bibfnamefont {H.}~\bibnamefont {Shtrikman}},\
  }\bibfield  {title} {\emph {\bibinfo {title} {Coupled electron-hole
  transport},\ }}\href {\doibase 10.1103/PhysRevLett.68.1196} {\bibfield
  {journal} {\bibinfo  {journal} {Physical Review Letters}\ }\textbf {\bibinfo
  {volume} {68}},\ \bibinfo {pages} {1196} (\bibinfo {year}
  {1992})}\BibitemShut {NoStop}%
\bibitem [{\citenamefont {Jauho}\ and\ \citenamefont
  {Smith}(1993)}]{jauho_coulomb_1993}%
  \BibitemOpen
  \bibfield  {author} {\bibinfo {author} {\bibfnamefont {A.-P.}\ \bibnamefont
  {Jauho}}\ and\ \bibinfo {author} {\bibfnamefont {H.}~\bibnamefont {Smith}},\
  }\bibfield  {title} {\emph {\bibinfo {title} {Coulomb drag between parallel
  two-dimensional electron systems},\ }}\href {\doibase
  10.1103/PhysRevB.47.4420} {\bibfield  {journal} {\bibinfo  {journal}
  {Physical Review B}\ }\textbf {\bibinfo {volume} {47}},\ \bibinfo {pages}
  {4420} (\bibinfo {year} {1993})}\BibitemShut {NoStop}%
\bibitem [{\citenamefont {Flensberg}\ and\ \citenamefont
  {Hu}(1994)}]{flensberg_coulomb_1994}%
  \BibitemOpen
  \bibfield  {author} {\bibinfo {author} {\bibfnamefont {K.}~\bibnamefont
  {Flensberg}}\ and\ \bibinfo {author} {\bibfnamefont {B.~Y.-K.}\ \bibnamefont
  {Hu}},\ }\bibfield  {title} {\emph {\bibinfo {title} {Coulomb {Drag} as a
  {Probe} of {Coupled} {Plasmon} {Modes} in {Parallel} {Quantum} {Wells}},\
  }}\href {\doibase 10.1103/PhysRevLett.73.3572} {\bibfield  {journal}
  {\bibinfo  {journal} {Physical Review Letters}\ }\textbf {\bibinfo {volume}
  {73}},\ \bibinfo {pages} {3572} (\bibinfo {year} {1994})}\BibitemShut
  {NoStop}%
\bibitem [{\citenamefont {Kamenev}\ and\ \citenamefont
  {Oreg}(1995)}]{kamenev_coulomb_1995}%
  \BibitemOpen
  \bibfield  {author} {\bibinfo {author} {\bibfnamefont {A.}~\bibnamefont
  {Kamenev}}\ and\ \bibinfo {author} {\bibfnamefont {Y.}~\bibnamefont {Oreg}},\
  }\bibfield  {title} {\emph {\bibinfo {title} {Coulomb drag in normal metals
  and superconductors: {Diagrammatic} approach},\ }}\href {\doibase
  10.1103/PhysRevB.52.7516} {\bibfield  {journal} {\bibinfo  {journal}
  {Physical Review B}\ }\textbf {\bibinfo {volume} {52}},\ \bibinfo {pages}
  {7516} (\bibinfo {year} {1995})}\BibitemShut {NoStop}%
\bibitem [{\citenamefont {Flensberg}\ \emph {et~al.}(1995)\citenamefont
  {Flensberg}, \citenamefont {Hu}, \citenamefont {Jauho},\ and\ \citenamefont
  {Kinaret}}]{flensberg_linear-response_1995}%
  \BibitemOpen
  \bibfield  {author} {\bibinfo {author} {\bibfnamefont {K.}~\bibnamefont
  {Flensberg}}, \bibinfo {author} {\bibfnamefont {B.~Y.-K.}\ \bibnamefont
  {Hu}}, \bibinfo {author} {\bibfnamefont {A.-P.}\ \bibnamefont {Jauho}}, \
  and\ \bibinfo {author} {\bibfnamefont {J.~M.}\ \bibnamefont {Kinaret}},\
  }\bibfield  {title} {\emph {\bibinfo {title} {Linear-response theory of
  {Coulomb} drag in coupled electron systems},\ }}\href {\doibase
  10.1103/PhysRevB.52.14761} {\bibfield  {journal} {\bibinfo  {journal}
  {Physical Review B}\ }\textbf {\bibinfo {volume} {52}},\ \bibinfo {pages}
  {14761} (\bibinfo {year} {1995})}\BibitemShut {NoStop}%
\bibitem [{\citenamefont {Nandi}\ \emph {et~al.}(2012)\citenamefont {Nandi},
  \citenamefont {Finck}, \citenamefont {Eisenstein}, \citenamefont {Pfeiffer},\
  and\ \citenamefont {West}}]{nandi_exciton_2012}%
  \BibitemOpen
  \bibfield  {author} {\bibinfo {author} {\bibfnamefont {D.}~\bibnamefont
  {Nandi}}, \bibinfo {author} {\bibfnamefont {A.~D.~K.}\ \bibnamefont {Finck}},
  \bibinfo {author} {\bibfnamefont {J.~P.}\ \bibnamefont {Eisenstein}},
  \bibinfo {author} {\bibfnamefont {L.~N.}\ \bibnamefont {Pfeiffer}}, \ and\
  \bibinfo {author} {\bibfnamefont {K.~W.}\ \bibnamefont {West}},\ }\bibfield
  {title} {\emph {{\bibinfo {title} {Exciton condensation
  and perfect {Coulomb} drag},\ }}}\href {\doibase 10.1038/nature11302}
  {\bibfield  {journal} {\bibinfo  {journal} {Nature}\ }\textbf {\bibinfo
  {volume} {488}},\ \bibinfo {pages} {481} (\bibinfo {year}
  {2012})}\BibitemShut {NoStop}%
\bibitem [{\citenamefont {Banerjee}\ \emph {et~al.}(2009)\citenamefont
  {Banerjee}, \citenamefont {Register}, \citenamefont {Tutuc}, \citenamefont
  {Reddy},\ and\ \citenamefont {MacDonald}}]{banerjee_bilayer_2009}%
  \BibitemOpen
  \bibfield  {author} {\bibinfo {author} {\bibfnamefont {S.~K.}\ \bibnamefont
  {Banerjee}}, \bibinfo {author} {\bibfnamefont {L.~F.}\ \bibnamefont
  {Register}}, \bibinfo {author} {\bibfnamefont {E.}~\bibnamefont {Tutuc}},
  \bibinfo {author} {\bibfnamefont {D.}~\bibnamefont {Reddy}}, \ and\ \bibinfo
  {author} {\bibfnamefont {A.~H.}\ \bibnamefont {MacDonald}},\ }\bibfield
  {title} {\emph {\bibinfo {title} {Bilayer {PseudoSpin} {Field}-{Effect}
  {Transistor} ({BiSFET}): {A} {Proposed} {New} {Logic} {Device}},\ }}\href
  {\doibase 10.1109/LED.2008.2009362} {\bibfield  {journal} {\bibinfo
  {journal} {IEEE Electron Device Letters}\ }\textbf {\bibinfo {volume} {30}},\
  \bibinfo {pages} {158} (\bibinfo {year} {2009})}\BibitemShut {NoStop}%
\bibitem [{\citenamefont {Mink}\ \emph {et~al.}(2012)\citenamefont {Mink},
  \citenamefont {Stoof}, \citenamefont {Duine}, \citenamefont {Polini},\ and\
  \citenamefont {Vignale}}]{mink_probing_2012}%
  \BibitemOpen
  \bibfield  {author} {\bibinfo {author} {\bibfnamefont {M.~P.}\ \bibnamefont
  {Mink}}, \bibinfo {author} {\bibfnamefont {H.~T.~C.}\ \bibnamefont {Stoof}},
  \bibinfo {author} {\bibfnamefont {R.~A.}\ \bibnamefont {Duine}}, \bibinfo
  {author} {\bibfnamefont {M.}~\bibnamefont {Polini}}, \ and\ \bibinfo {author}
  {\bibfnamefont {G.}~\bibnamefont {Vignale}},\ }\bibfield  {title} {\emph
  {\bibinfo {title} {Probing the {Topological} {Exciton} {Condensate} via
  {Coulomb} {Drag}},\ }}\href {\doibase 10.1103/PhysRevLett.108.186402}
  {\bibfield  {journal} {\bibinfo  {journal} {Physical Review Letters}\
  }\textbf {\bibinfo {volume} {108}},\ \bibinfo {pages} {186402} (\bibinfo
  {year} {2012})}\BibitemShut {NoStop}%
\bibitem [{\citenamefont {Efimkin}\ and\ \citenamefont
  {Galitski}(2016)}]{efimkin_anomalous_2016}%
  \BibitemOpen
  \bibfield  {author} {\bibinfo {author} {\bibfnamefont {D.~K.}\ \bibnamefont
  {Efimkin}}\ and\ \bibinfo {author} {\bibfnamefont {V.}~\bibnamefont
  {Galitski}},\ }\bibfield  {title} {\emph {\bibinfo {title} {Anomalous
  {Coulomb} {Drag} in {Electron}-{Hole} {Bilayers} due to the {Formation} of
  {Excitons}},\ }}\href {\doibase 10.1103/PhysRevLett.116.046801} {\bibfield
  {journal} {\bibinfo  {journal} {Physical Review Letters}\ }\textbf {\bibinfo
  {volume} {116}},\ \bibinfo {pages} {046801} (\bibinfo {year}
  {2016})}\BibitemShut {NoStop}%
\bibitem [{\citenamefont {Kim}\ \emph {et~al.}(2011)\citenamefont {Kim},
  \citenamefont {Jo}, \citenamefont {Nah}, \citenamefont {Yao}, \citenamefont
  {Banerjee},\ and\ \citenamefont {Tutuc}}]{kim_coulomb_2011}%
  \BibitemOpen
  \bibfield  {author} {\bibinfo {author} {\bibfnamefont {S.}~\bibnamefont
  {Kim}}, \bibinfo {author} {\bibfnamefont {I.}~\bibnamefont {Jo}}, \bibinfo
  {author} {\bibfnamefont {J.}~\bibnamefont {Nah}}, \bibinfo {author}
  {\bibfnamefont {Z.}~\bibnamefont {Yao}}, \bibinfo {author} {\bibfnamefont
  {S.~K.}\ \bibnamefont {Banerjee}}, \ and\ \bibinfo {author} {\bibfnamefont
  {E.}~\bibnamefont {Tutuc}},\ }\bibfield  {title} {\emph {\bibinfo {title}
  {Coulomb drag of massless fermions in graphene},\ }}\href {\doibase
  10.1103/PhysRevB.83.161401} {\bibfield  {journal} {\bibinfo  {journal}
  {Physical Review B}\ }\textbf {\bibinfo {volume} {83}},\ \bibinfo {pages}
  {161401} (\bibinfo {year} {2011})}\BibitemShut {NoStop}%
\bibitem [{\citenamefont {Kim}\ and\ \citenamefont
  {Tutuc}(2012)}]{kim_coulomb_2012}%
  \BibitemOpen
  \bibfield  {author} {\bibinfo {author} {\bibfnamefont {S.}~\bibnamefont
  {Kim}}\ and\ \bibinfo {author} {\bibfnamefont {E.}~\bibnamefont {Tutuc}},\
  }\bibfield  {title} {\emph {\bibinfo {title} {Coulomb drag and
  magnetotransport in graphene double layers},\ }}\href {\doibase
  10.1016/j.ssc.2012.04.032} {\bibfield  {journal} {\bibinfo  {journal} {Solid
  State Communications}\ }\bibinfo {series} {Exploring {Graphene}, {Recent}
  {Research} {Advances}},\ \textbf {\bibinfo {volume} {152}},\ \bibinfo {pages}
  {1283} (\bibinfo {year} {2012})}\BibitemShut {NoStop}%
\bibitem [{\citenamefont {Gorbachev}\ \emph {et~al.}(2012)\citenamefont
  {Gorbachev}, \citenamefont {Geim}, \citenamefont {Katsnelson}, \citenamefont
  {Novoselov}, \citenamefont {Tudorovskiy}, \citenamefont {Grigorieva},
  \citenamefont {MacDonald}, \citenamefont {Morozov}, \citenamefont {Watanabe},
  \citenamefont {Taniguchi},\ and\ \citenamefont
  {Ponomarenko}}]{gorbachev_strong_2012}%
  \BibitemOpen
  \bibfield  {author} {\bibinfo {author} {\bibfnamefont {R.~V.}\ \bibnamefont
  {Gorbachev}}, \bibinfo {author} {\bibfnamefont {A.~K.}\ \bibnamefont {Geim}},
  \bibinfo {author} {\bibfnamefont {M.~I.}\ \bibnamefont {Katsnelson}},
  \bibinfo {author} {\bibfnamefont {K.~S.}\ \bibnamefont {Novoselov}}, \bibinfo
  {author} {\bibfnamefont {T.}~\bibnamefont {Tudorovskiy}}, \bibinfo {author}
  {\bibfnamefont {I.~V.}\ \bibnamefont {Grigorieva}}, \bibinfo {author}
  {\bibfnamefont {A.~H.}\ \bibnamefont {MacDonald}}, \bibinfo {author}
  {\bibfnamefont {S.~V.}\ \bibnamefont {Morozov}}, \bibinfo {author}
  {\bibfnamefont {K.}~\bibnamefont {Watanabe}}, \bibinfo {author}
  {\bibfnamefont {T.}~\bibnamefont {Taniguchi}}, \ and\ \bibinfo {author}
  {\bibfnamefont {L.~A.}\ \bibnamefont {Ponomarenko}},\ }\bibfield  {title}
  {\emph {{\bibinfo {title} {Strong {Coulomb} drag and
  broken symmetry in double-layer graphene},\ }}}\href {\doibase
  10.1038/nphys2441} {\bibfield  {journal} {\bibinfo  {journal} {Nature
  Physics}\ }\textbf {\bibinfo {volume} {8}},\ \bibinfo {pages} {896} (\bibinfo
  {year} {2012})}\BibitemShut {NoStop}%
\bibitem [{\citenamefont {Titov}\ \emph {et~al.}(2013)\citenamefont {Titov},
  \citenamefont {Gorbachev}, \citenamefont {Narozhny}, \citenamefont
  {Tudorovskiy}, \citenamefont {Schütt}, \citenamefont {Ostrovsky},
  \citenamefont {Gornyi}, \citenamefont {Mirlin}, \citenamefont {Katsnelson},
  \citenamefont {Novoselov}, \citenamefont {Geim},\ and\ \citenamefont
  {Ponomarenko}}]{titov_giant_2013}%
  \BibitemOpen
  \bibfield  {author} {\bibinfo {author} {\bibfnamefont {M.}~\bibnamefont
  {Titov}}, \bibinfo {author} {\bibfnamefont {R.~V.}\ \bibnamefont
  {Gorbachev}}, \bibinfo {author} {\bibfnamefont {B.~N.}\ \bibnamefont
  {Narozhny}}, \bibinfo {author} {\bibfnamefont {T.}~\bibnamefont
  {Tudorovskiy}}, \bibinfo {author} {\bibfnamefont {M.}~\bibnamefont
  {Schütt}}, \bibinfo {author} {\bibfnamefont {P.~M.}\ \bibnamefont
  {Ostrovsky}}, \bibinfo {author} {\bibfnamefont {I.~V.}\ \bibnamefont
  {Gornyi}}, \bibinfo {author} {\bibfnamefont {A.~D.}\ \bibnamefont {Mirlin}},
  \bibinfo {author} {\bibfnamefont {M.~I.}\ \bibnamefont {Katsnelson}},
  \bibinfo {author} {\bibfnamefont {K.~S.}\ \bibnamefont {Novoselov}}, \bibinfo
  {author} {\bibfnamefont {A.~K.}\ \bibnamefont {Geim}}, \ and\ \bibinfo
  {author} {\bibfnamefont {L.~A.}\ \bibnamefont {Ponomarenko}},\ }\bibfield
  {title} {\emph {\bibinfo {title} {Giant {Magnetodrag} in {Graphene} at
  {Charge} {Neutrality}},\ }}\href {\doibase 10.1103/PhysRevLett.111.166601}
  {\bibfield  {journal} {\bibinfo  {journal} {Physical Review Letters}\
  }\textbf {\bibinfo {volume} {111}},\ \bibinfo {pages} {166601} (\bibinfo
  {year} {2013})}\BibitemShut {NoStop}%
\bibitem [{\citenamefont {Li}\ \emph {et~al.}(2016{\natexlab{a}})\citenamefont
  {Li}, \citenamefont {Taniguchi}, \citenamefont {Watanabe}, \citenamefont
  {Hone},\ and\ \citenamefont {Dean}}]{li_excitonic_2016}%
  \BibitemOpen
  \bibfield  {author} {\bibinfo {author} {\bibfnamefont {J.~I.~A.}\
  \bibnamefont {Li}}, \bibinfo {author} {\bibfnamefont {T.}~\bibnamefont
  {Taniguchi}}, \bibinfo {author} {\bibfnamefont {K.}~\bibnamefont {Watanabe}},
  \bibinfo {author} {\bibfnamefont {J.}~\bibnamefont {Hone}}, \ and\ \bibinfo
  {author} {\bibfnamefont {C.~R.}\ \bibnamefont {Dean}},\ }\bibfield  {title}
  {\emph {\bibinfo {title} {Excitonic superfluid phase in {Double} {Bilayer}
  {Graphene}},\ }}\href {http://arxiv.org/abs/1608.05846} {\bibfield  {journal}
  {\bibinfo  {journal} {arXiv:1608.05846 [cond-mat]}\ } (\bibinfo {year}
  {2016}{\natexlab{a}})},\ \BibitemShut
  {NoStop}%
\bibitem [{\citenamefont {Lee}\ \emph {et~al.}(2016)\citenamefont {Lee},
  \citenamefont {Xue}, \citenamefont {Dillen}, \citenamefont {Watanabe},
  \citenamefont {Taniguchi},\ and\ \citenamefont {Tutuc}}]{lee_giant_2016}%
  \BibitemOpen
  \bibfield  {author} {\bibinfo {author} {\bibfnamefont {K.}~\bibnamefont
  {Lee}}, \bibinfo {author} {\bibfnamefont {J.}~\bibnamefont {Xue}}, \bibinfo
  {author} {\bibfnamefont {D.~C.}\ \bibnamefont {Dillen}}, \bibinfo {author}
  {\bibfnamefont {K.}~\bibnamefont {Watanabe}}, \bibinfo {author}
  {\bibfnamefont {T.}~\bibnamefont {Taniguchi}}, \ and\ \bibinfo {author}
  {\bibfnamefont {E.}~\bibnamefont {Tutuc}},\ }\bibfield  {title} {\emph
  {\bibinfo {title} {Giant {Frictional} {Drag} in {Double} {Bilayer} {Graphene}
  {Heterostructures}},\ }}\href {\doibase 10.1103/PhysRevLett.117.046803}
  {\bibfield  {journal} {\bibinfo  {journal} {Physical Review Letters}\
  }\textbf {\bibinfo {volume} {117}},\ \bibinfo {pages} {046803} (\bibinfo
  {year} {2016})}\BibitemShut {NoStop}%
\bibitem [{\citenamefont {Li}\ \emph {et~al.}(2016{\natexlab{b}})\citenamefont
  {Li}, \citenamefont {Taniguchi}, \citenamefont {Watanabe}, \citenamefont
  {Hone}, \citenamefont {Levchenko},\ and\ \citenamefont
  {Dean}}]{li_negative_2016}%
  \BibitemOpen
  \bibfield  {author} {\bibinfo {author} {\bibfnamefont {J.}~\bibnamefont
  {Li}}, \bibinfo {author} {\bibfnamefont {T.}~\bibnamefont {Taniguchi}},
  \bibinfo {author} {\bibfnamefont {K.}~\bibnamefont {Watanabe}}, \bibinfo
  {author} {\bibfnamefont {J.}~\bibnamefont {Hone}}, \bibinfo {author}
  {\bibfnamefont {A.}~\bibnamefont {Levchenko}}, \ and\ \bibinfo {author}
  {\bibfnamefont {C.}~\bibnamefont {Dean}},\ }\bibfield  {title} {\emph
  {\bibinfo {title} {Negative {Coulomb} {Drag} in {Double} {Bilayer}
  {Graphene}},\ }}\href {\doibase 10.1103/PhysRevLett.117.046802} {\bibfield
  {journal} {\bibinfo  {journal} {Physical Review Letters}\ }\textbf {\bibinfo
  {volume} {117}},\ \bibinfo {pages} {046802} (\bibinfo {year}
  {2016}{\natexlab{b}})}\BibitemShut {NoStop}%
\bibitem [{\citenamefont {Liu}\ \emph {et~al.}(2016)\citenamefont {Liu},
  \citenamefont {Watanabe}, \citenamefont {Taniguchi}, \citenamefont
  {Halperin},\ and\ \citenamefont {Kim}}]{liu_quantum_2016}%
  \BibitemOpen
  \bibfield  {author} {\bibinfo {author} {\bibfnamefont {X.}~\bibnamefont
  {Liu}}, \bibinfo {author} {\bibfnamefont {K.}~\bibnamefont {Watanabe}},
  \bibinfo {author} {\bibfnamefont {T.}~\bibnamefont {Taniguchi}}, \bibinfo
  {author} {\bibfnamefont {B.~I.}\ \bibnamefont {Halperin}}, \ and\ \bibinfo
  {author} {\bibfnamefont {P.}~\bibnamefont {Kim}},\ }\bibfield  {title} {\emph
  {\bibinfo {title} {Quantum {Hall} {Drag} of {Exciton} {Superfluid} in
  {Graphene}},\ }}\href {http://arxiv.org/abs/1608.03726} {\bibfield  {journal}
  {\bibinfo  {journal} {arXiv:1608.03726 [cond-mat]}\ } (\bibinfo {year}
  {2016})},\ \BibitemShut {NoStop}%
\bibitem [{\citenamefont {Martin}\ \emph {et~al.}(2008)\citenamefont {Martin},
  \citenamefont {Akerman}, \citenamefont {Ulbricht}, \citenamefont {Lohmann},
  \citenamefont {Smet}, \citenamefont {von Klitzing},\ and\ \citenamefont
  {Yacoby}}]{martin_observation_2008}%
  \BibitemOpen
  \bibfield  {author} {\bibinfo {author} {\bibfnamefont {J.}~\bibnamefont
  {Martin}}, \bibinfo {author} {\bibfnamefont {N.}~\bibnamefont {Akerman}},
  \bibinfo {author} {\bibfnamefont {G.}~\bibnamefont {Ulbricht}}, \bibinfo
  {author} {\bibfnamefont {T.}~\bibnamefont {Lohmann}}, \bibinfo {author}
  {\bibfnamefont {J.~H.}\ \bibnamefont {Smet}}, \bibinfo {author}
  {\bibfnamefont {K.}~\bibnamefont {von Klitzing}}, \ and\ \bibinfo {author}
  {\bibfnamefont {A.}~\bibnamefont {Yacoby}},\ }\bibfield  {title} {\emph
  {{\bibinfo {title} {Observation of electron–hole
  puddles in graphene using a scanning single-electron transistor},\ }}}\href
  {\doibase 10.1038/nphys781} {\bibfield  {journal} {\bibinfo  {journal}
  {Nature Physics}\ }\textbf {\bibinfo {volume} {4}},\ \bibinfo {pages} {144}
  (\bibinfo {year} {2008})}\BibitemShut {NoStop}%
\bibitem [{\citenamefont {Adam}\ \emph {et~al.}(2007)\citenamefont {Adam},
  \citenamefont {Hwang}, \citenamefont {Galitski},\ and\ \citenamefont
  {Sarma}}]{adam_self-consistent_2007}%
  \BibitemOpen
  \bibfield  {author} {\bibinfo {author} {\bibfnamefont {S.}~\bibnamefont
  {Adam}}, \bibinfo {author} {\bibfnamefont {E.~H.}\ \bibnamefont {Hwang}},
  \bibinfo {author} {\bibfnamefont {V.~M.}\ \bibnamefont {Galitski}}, \ and\
  \bibinfo {author} {\bibfnamefont {S.~D.}\ \bibnamefont {Sarma}},\ }\bibfield
  {title} {\emph {{\bibinfo {title} {A self-consistent
  theory for graphene transport},\ }}}\href {\doibase 10.1073/pnas.0704772104}
  {\bibfield  {journal} {\bibinfo  {journal} {Proceedings of the National
  Academy of Sciences}\ }\textbf {\bibinfo {volume} {104}},\ \bibinfo {pages}
  {18392} (\bibinfo {year} {2007})}\BibitemShut {NoStop}%
\bibitem [{\citenamefont {Gibertini}\ \emph {et~al.}(2012)\citenamefont
  {Gibertini}, \citenamefont {Tomadin}, \citenamefont {Guinea}, \citenamefont
  {Katsnelson},\ and\ \citenamefont {Polini}}]{gibertini_electron-hole_2012}%
  \BibitemOpen
  \bibfield  {author} {\bibinfo {author} {\bibfnamefont {M.}~\bibnamefont
  {Gibertini}}, \bibinfo {author} {\bibfnamefont {A.}~\bibnamefont {Tomadin}},
  \bibinfo {author} {\bibfnamefont {F.}~\bibnamefont {Guinea}}, \bibinfo
  {author} {\bibfnamefont {M.~I.}\ \bibnamefont {Katsnelson}}, \ and\ \bibinfo
  {author} {\bibfnamefont {M.}~\bibnamefont {Polini}},\ }\bibfield  {title}
  {\emph {\bibinfo {title} {Electron-hole puddles in the absence of charged
  impurities},\ }}\href {\doibase 10.1103/PhysRevB.85.201405} {\bibfield
  {journal} {\bibinfo  {journal} {Physical Review B}\ }\textbf {\bibinfo
  {volume} {85}},\ \bibinfo {pages} {201405} (\bibinfo {year}
  {2012})}\BibitemShut {NoStop}%
\bibitem [{\citenamefont {Landauer}(1978)}]{landauer_electrical_1978}%
  \BibitemOpen
  \bibfield  {author} {\bibinfo {author} {\bibfnamefont {R.}~\bibnamefont
  {Landauer}},\ }in\ \href {\doibase 10.1063/1.31150} {\emph {\bibinfo
  {booktitle} {{AIP} {Conference} {Proceedings}}}},\ Vol.~\bibinfo {volume}
  {40}\ (\bibinfo  {publisher} {AIP Publishing},\ \bibinfo {year} {1978})\ pp.\
  \bibinfo {pages} {2--45}\BibitemShut {NoStop}%
\bibitem [{\citenamefont {Stroud}(1998)}]{stroud_effective_1998}%
  \BibitemOpen
  \bibfield  {author} {\bibinfo {author} {\bibfnamefont {D.}~\bibnamefont
  {Stroud}},\ }\bibfield  {title} {\emph {\bibinfo {title} {The effective
  medium approximations: {Some} recent developments},\ }}\href {\doibase
  10.1006/spmi.1997.0524} {\bibfield  {journal} {\bibinfo  {journal}
  {Superlattices and Microstructures}\ }\textbf {\bibinfo {volume} {23}},\
  \bibinfo {pages} {567} (\bibinfo {year} {1998})}\BibitemShut {NoStop}%
\bibitem [{\citenamefont {Choy}(2016)}]{choy_effective_2016}%
  \BibitemOpen
  \bibfield  {author} {\bibinfo {author} {\bibfnamefont {T.~C.}\ \bibnamefont
  {Choy}},\ }\href@noop {} {{\emph {\bibinfo {title}
  {Effective {Medium} {Theory}: {Principles} and {Applications}}}}}\ (\bibinfo
  {publisher} {Oxford University Press},\ \bibinfo {year} {2016})\BibitemShut
  {NoStop}%
\bibitem [{\citenamefont {Seamons}\ \emph {et~al.}(2009)\citenamefont
  {Seamons}, \citenamefont {Morath}, \citenamefont {Reno},\ and\ \citenamefont
  {Lilly}}]{seamons_coulomb_2009}%
  \BibitemOpen
  \bibfield  {author} {\bibinfo {author} {\bibfnamefont {J.~A.}\ \bibnamefont
  {Seamons}}, \bibinfo {author} {\bibfnamefont {C.~P.}\ \bibnamefont {Morath}},
  \bibinfo {author} {\bibfnamefont {J.~L.}\ \bibnamefont {Reno}}, \ and\
  \bibinfo {author} {\bibfnamefont {M.~P.}\ \bibnamefont {Lilly}},\ }\bibfield
  {title} {\emph {\bibinfo {title} {Coulomb {Drag} in the {Exciton} {Regime} in
  {Electron}-{Hole} {Bilayers}},\ }}\href {\doibase
  10.1103/PhysRevLett.102.026804} {\bibfield  {journal} {\bibinfo  {journal}
  {Physical Review Letters}\ }\textbf {\bibinfo {volume} {102}},\ \bibinfo
  {pages} {026804} (\bibinfo {year} {2009})}\BibitemShut {NoStop}%
\bibitem [{\citenamefont {Vignale}\ and\ \citenamefont
  {MacDonald}(1996)}]{vignale_drag_1996}%
  \BibitemOpen
  \bibfield  {author} {\bibinfo {author} {\bibfnamefont {G.}~\bibnamefont
  {Vignale}}\ and\ \bibinfo {author} {\bibfnamefont {A.~H.}\ \bibnamefont
  {MacDonald}},\ }\bibfield  {title} {\emph {\bibinfo {title} {Drag in {Paired}
  {Electron}-{Hole} {Layers}},\ }}\href {\doibase 10.1103/PhysRevLett.76.2786}
  {\bibfield  {journal} {\bibinfo  {journal} {Physical Review Letters}\
  }\textbf {\bibinfo {volume} {76}},\ \bibinfo {pages} {2786} (\bibinfo {year}
  {1996})}\BibitemShut {NoStop}%
\bibitem [{\citenamefont {Song}\ and\ \citenamefont
  {Levitov}(2012)}]{song_energy-driven_2012}%
  \BibitemOpen
  \bibfield  {author} {\bibinfo {author} {\bibfnamefont {J.~C.~W.}\
  \bibnamefont {Song}}\ and\ \bibinfo {author} {\bibfnamefont {L.~S.}\
  \bibnamefont {Levitov}},\ }\bibfield  {title} {\emph {\bibinfo {title}
  {Energy-{Driven} {Drag} at {Charge} {Neutrality} in {Graphene}},\ }}\href
  {\doibase 10.1103/PhysRevLett.109.236602} {\bibfield  {journal} {\bibinfo
  {journal} {Physical Review Letters}\ }\textbf {\bibinfo {volume} {109}},\
  \bibinfo {pages} {236602} (\bibinfo {year} {2012})}\BibitemShut {NoStop}%
\bibitem [{\citenamefont {Apalkov}\ and\ \citenamefont
  {Raikh}(2005)}]{apalkov_effective_2005}%
  \BibitemOpen
  \bibfield  {author} {\bibinfo {author} {\bibfnamefont {V.~M.}\ \bibnamefont
  {Apalkov}}\ and\ \bibinfo {author} {\bibfnamefont {M.~E.}\ \bibnamefont
  {Raikh}},\ }\bibfield  {title} {\emph {\bibinfo {title} {Effective drag
  between strongly inhomogeneous layers: {Exact} results and applications},\
  }}\href {\doibase 10.1103/PhysRevB.71.245109} {\bibfield  {journal} {\bibinfo
   {journal} {Physical Review B}\ }\textbf {\bibinfo {volume} {71}},\ \bibinfo
  {pages} {245109} (\bibinfo {year} {2005})}\BibitemShut {NoStop}%
\bibitem [{\citenamefont {Narozhny}\ \emph {et~al.}(2012)\citenamefont
  {Narozhny}, \citenamefont {Titov}, \citenamefont {Gornyi},\ and\
  \citenamefont {Ostrovsky}}]{narozhny_coulomb_2012}%
  \BibitemOpen
  \bibfield  {author} {\bibinfo {author} {\bibfnamefont {B.~N.}\ \bibnamefont
  {Narozhny}}, \bibinfo {author} {\bibfnamefont {M.}~\bibnamefont {Titov}},
  \bibinfo {author} {\bibfnamefont {I.~V.}\ \bibnamefont {Gornyi}}, \ and\
  \bibinfo {author} {\bibfnamefont {P.~M.}\ \bibnamefont {Ostrovsky}},\
  }\bibfield  {title} {\emph {\bibinfo {title} {Coulomb drag in graphene:
  {Perturbation} theory},\ }}\href {\doibase 10.1103/PhysRevB.85.195421}
  {\bibfield  {journal} {\bibinfo  {journal} {Physical Review B}\ }\textbf
  {\bibinfo {volume} {85}},\ \bibinfo {pages} {195421} (\bibinfo {year}
  {2012})}\BibitemShut {NoStop}%
\bibitem [{\citenamefont {Min}\ \emph {et~al.}(2008)\citenamefont {Min},
  \citenamefont {Bistritzer}, \citenamefont {Su},\ and\ \citenamefont
  {MacDonald}}]{min_room-temperature_2008}%
  \BibitemOpen
  \bibfield  {author} {\bibinfo {author} {\bibfnamefont {H.}~\bibnamefont
  {Min}}, \bibinfo {author} {\bibfnamefont {R.}~\bibnamefont {Bistritzer}},
  \bibinfo {author} {\bibfnamefont {J.-J.}\ \bibnamefont {Su}}, \ and\ \bibinfo
  {author} {\bibfnamefont {A.~H.}\ \bibnamefont {MacDonald}},\ }\bibfield
  {title} {\emph {\bibinfo {title} {Room-temperature superfluidity in graphene
  bilayers},\ }}\href {\doibase 10.1103/PhysRevB.78.121401} {\bibfield
  {journal} {\bibinfo  {journal} {Physical Review B}\ }\textbf {\bibinfo
  {volume} {78}},\ \bibinfo {pages} {121401} (\bibinfo {year}
  {2008})}\BibitemShut {NoStop}%
\bibitem [{\citenamefont {Lozovik}\ \emph {et~al.}(2012)\citenamefont
  {Lozovik}, \citenamefont {Ogarkov},\ and\ \citenamefont
  {Sokolik}}]{lozovik_condensation_2012}%
  \BibitemOpen
  \bibfield  {author} {\bibinfo {author} {\bibfnamefont {Y.~E.}\ \bibnamefont
  {Lozovik}}, \bibinfo {author} {\bibfnamefont {S.~L.}\ \bibnamefont
  {Ogarkov}}, \ and\ \bibinfo {author} {\bibfnamefont {A.~A.}\ \bibnamefont
  {Sokolik}},\ }\bibfield  {title} {\emph {\bibinfo {title} {Condensation of
  electron-hole pairs in a two-layer graphene system: {Correlation} effects},\
  }}\href {\doibase 10.1103/PhysRevB.86.045429} {\bibfield  {journal} {\bibinfo
   {journal} {Physical Review B}\ }\textbf {\bibinfo {volume} {86}},\ \bibinfo
  {pages} {045429} (\bibinfo {year} {2012})}\BibitemShut {NoStop}%
\bibitem [{\citenamefont {de~Cumis}\ \emph {et~al.}(2016)\citenamefont
  {de~Cumis}, \citenamefont {Waldie}, \citenamefont {Croxall}, \citenamefont
  {Taneja}, \citenamefont {Llandro}, \citenamefont {Farrer}, \citenamefont
  {Beere},\ and\ \citenamefont {Ritchie}}]{de_cumis_complete_2016}%
  \BibitemOpen
  \bibfield  {author} {\bibinfo {author} {\bibfnamefont {U.~S.}\ \bibnamefont
  {de~Cumis}}, \bibinfo {author} {\bibfnamefont {J.}~\bibnamefont {Waldie}},
  \bibinfo {author} {\bibfnamefont {A.~F.}\ \bibnamefont {Croxall}}, \bibinfo
  {author} {\bibfnamefont {D.}~\bibnamefont {Taneja}}, \bibinfo {author}
  {\bibfnamefont {J.}~\bibnamefont {Llandro}}, \bibinfo {author} {\bibfnamefont
  {I.}~\bibnamefont {Farrer}}, \bibinfo {author} {\bibfnamefont {H.~E.}\
  \bibnamefont {Beere}}, \ and\ \bibinfo {author} {\bibfnamefont {D.~A.}\
  \bibnamefont {Ritchie}},\ }\bibfield  {title} {\emph {\bibinfo {title} {A
  complete laboratory for transport studies of electron-hole interactions in
  {GaAs}/{AlGaAs} systems},\ }}\href {http://arxiv.org/abs/1611.08816}
  {\bibfield  {journal} {\bibinfo  {journal} {arXiv:1611.08816 [cond-mat]}\ }
  (\bibinfo {year} {2016})},\ \BibitemShut
  {NoStop}%
\bibitem [{\citenamefont {Tse}\ \emph {et~al.}(2007)\citenamefont {Tse},
  \citenamefont {Hu},\ and\ \citenamefont {Das~Sarma}}]{tse_theory_2007}%
  \BibitemOpen
  \bibfield  {author} {\bibinfo {author} {\bibfnamefont {W.-K.}\ \bibnamefont
  {Tse}}, \bibinfo {author} {\bibfnamefont {B.~Y.-K.}\ \bibnamefont {Hu}}, \
  and\ \bibinfo {author} {\bibfnamefont {S.}~\bibnamefont {Das~Sarma}},\
  }\bibfield  {title} {\emph {\bibinfo {title} {Theory of {Coulomb} drag in
  graphene},\ }}\href {\doibase 10.1103/PhysRevB.76.081401} {\bibfield
  {journal} {\bibinfo  {journal} {Physical Review B}\ }\textbf {\bibinfo
  {volume} {76}},\ \bibinfo {pages} {081401} (\bibinfo {year}
  {2007})}\BibitemShut {NoStop}%
\bibitem [{\citenamefont {Peres}\ \emph {et~al.}(2011)\citenamefont {Peres},
  \citenamefont {Santos},\ and\ \citenamefont {Neto}}]{peres_coulomb_2011}%
  \BibitemOpen
  \bibfield  {author} {\bibinfo {author} {\bibfnamefont {N.~M.~R.}\
  \bibnamefont {Peres}}, \bibinfo {author} {\bibfnamefont {J.~M. B. L.~d.}\
  \bibnamefont {Santos}}, \ and\ \bibinfo {author} {\bibfnamefont {A.~H.~C.}\
  \bibnamefont {Neto}},\ }\bibfield  {title} {\emph {{\bibinfo {title} {Coulomb drag and high-resistivity behavior in
  double-layer graphene},\ }}}\href {\doibase 10.1209/0295-5075/95/18001}
  {\bibfield  {journal} {\bibinfo  {journal} {EPL (Europhysics Letters)}\
  }\textbf {\bibinfo {volume} {95}},\ \bibinfo {pages} {18001} (\bibinfo {year}
  {2011})}\BibitemShut {NoStop}%
\bibitem [{\citenamefont {Carrega}\ \emph {et~al.}(2012)\citenamefont
  {Carrega}, \citenamefont {Tudorovskiy}, \citenamefont {Principi},
  \citenamefont {Katsnelson},\ and\ \citenamefont
  {Polini}}]{carrega_theory_2012}%
  \BibitemOpen
  \bibfield  {author} {\bibinfo {author} {\bibfnamefont {M.}~\bibnamefont
  {Carrega}}, \bibinfo {author} {\bibfnamefont {T.}~\bibnamefont
  {Tudorovskiy}}, \bibinfo {author} {\bibfnamefont {A.}~\bibnamefont
  {Principi}}, \bibinfo {author} {\bibfnamefont {M.~I.}\ \bibnamefont
  {Katsnelson}}, \ and\ \bibinfo {author} {\bibfnamefont {M.}~\bibnamefont
  {Polini}},\ }\bibfield  {title} {\emph {{\bibinfo {title}
  {Theory of {Coulomb} drag for massless {Dirac} fermions},\ }}}\href {\doibase
  10.1088/1367-2630/14/6/063033} {\bibfield  {journal} {\bibinfo  {journal}
  {New Journal of Physics}\ }\textbf {\bibinfo {volume} {14}},\ \bibinfo
  {pages} {063033} (\bibinfo {year} {2012})}\BibitemShut {NoStop}%
\bibitem [{\citenamefont {Amorim}\ and\ \citenamefont
  {Peres}(2012)}]{amorim_coulomb_2012}%
  \BibitemOpen
  \bibfield  {author} {\bibinfo {author} {\bibfnamefont {B.}~\bibnamefont
  {Amorim}}\ and\ \bibinfo {author} {\bibfnamefont {N.~M.~R.}\ \bibnamefont
  {Peres}},\ }\bibfield  {title} {\emph {{\bibinfo {title}
  {On {Coulomb} drag in double layer systems},\ }}}\href {\doibase
  10.1088/0953-8984/24/33/335602} {\bibfield  {journal} {\bibinfo  {journal}
  {Journal of Physics: Condensed Matter}\ }\textbf {\bibinfo {volume} {24}},\
  \bibinfo {pages} {335602} (\bibinfo {year} {2012})}\BibitemShut {NoStop}%
\bibitem [{\citenamefont {Ramezanali}\ \emph {et~al.}(2009)\citenamefont
  {Ramezanali}, \citenamefont {Vazifeh}, \citenamefont {Asgari}, \citenamefont
  {Polini},\ and\ \citenamefont
  {MacDonald}}]{ramezanali_finite-temperature_2009}%
  \BibitemOpen
  \bibfield  {author} {\bibinfo {author} {\bibfnamefont {M.~R.}\ \bibnamefont
  {Ramezanali}}, \bibinfo {author} {\bibfnamefont {M.~M.}\ \bibnamefont
  {Vazifeh}}, \bibinfo {author} {\bibfnamefont {R.}~\bibnamefont {Asgari}},
  \bibinfo {author} {\bibfnamefont {M.}~\bibnamefont {Polini}}, \ and\ \bibinfo
  {author} {\bibfnamefont {A.~H.}\ \bibnamefont {MacDonald}},\ }\bibfield
  {title} {\emph {\bibinfo {title} {Finite-temperature screening and the
  specific heat of doped graphene sheets},\ }}\href {\doibase
  10.1088/1751-8113/42/21/214015} {\bibfield  {journal} {\bibinfo  {journal}
  {Journal of Physics A: Mathematical and Theoretical}\ }\textbf {\bibinfo
  {volume} {42}},\ \bibinfo {pages} {214015} (\bibinfo {year}
  {2009})}\BibitemShut {NoStop}%
\bibitem [{\citenamefont {Badalyan}\ and\ \citenamefont
  {Peeters}(2012)}]{badalyan_enhancement_2012}%
  \BibitemOpen
  \bibfield  {author} {\bibinfo {author} {\bibfnamefont {S.~M.}\ \bibnamefont
  {Badalyan}}\ and\ \bibinfo {author} {\bibfnamefont {F.~M.}\ \bibnamefont
  {Peeters}},\ }\bibfield  {title} {\emph {\bibinfo {title} {Enhancement of
  {Coulomb} drag in double-layer graphene structures by plasmons and dielectric
  background inhomogeneity},\ }}\href {\doibase 10.1103/PhysRevB.86.121405}
  {\bibfield  {journal} {\bibinfo  {journal} {Physical Review B}\ }\textbf
  {\bibinfo {volume} {86}},\ \bibinfo {pages} {121405} (\bibinfo {year}
  {2012})}\BibitemShut {NoStop}%
\bibitem [{\citenamefont {Amorim}\ \emph {et~al.}(2012)\citenamefont {Amorim},
  \citenamefont {Schiefele}, \citenamefont {Sols},\ and\ \citenamefont
  {Guinea}}]{amorim_coulomb_2012-1}%
  \BibitemOpen
  \bibfield  {author} {\bibinfo {author} {\bibfnamefont {B.}~\bibnamefont
  {Amorim}}, \bibinfo {author} {\bibfnamefont {J.}~\bibnamefont {Schiefele}},
  \bibinfo {author} {\bibfnamefont {F.}~\bibnamefont {Sols}}, \ and\ \bibinfo
  {author} {\bibfnamefont {F.}~\bibnamefont {Guinea}},\ }\bibfield  {title}
  {\emph {\bibinfo {title} {Coulomb drag in
  graphene{\textbackslash}char21\{\}boron nitride heterostructures: {Effect} of
  virtual phonon exchange},\ }}\href {\doibase 10.1103/PhysRevB.86.125448}
  {\bibfield  {journal} {\bibinfo  {journal} {Physical Review B}\ }\textbf
  {\bibinfo {volume} {86}},\ \bibinfo {pages} {125448} (\bibinfo {year}
  {2012})}\BibitemShut {NoStop}%
\bibitem [{\citenamefont {Lung}\ and\ \citenamefont
  {Marinescu}(2011)}]{lung_thermoelectric_2011}%
  \BibitemOpen
  \bibfield  {author} {\bibinfo {author} {\bibfnamefont {F.}~\bibnamefont
  {Lung}}\ and\ \bibinfo {author} {\bibfnamefont {D.~C.}\ \bibnamefont
  {Marinescu}},\ }\bibfield  {title} {\emph {\bibinfo {title} {Thermoelectric
  effect in a bi-layer system},\ }}\href {\doibase 10.1016/j.physe.2011.06.003}
  {\bibfield  {journal} {\bibinfo  {journal} {Physica E: Low-dimensional
  Systems and Nanostructures}\ }\textbf {\bibinfo {volume} {43}},\ \bibinfo
  {pages} {1769} (\bibinfo {year} {2011})}\BibitemShut {NoStop}%
\bibitem [{\citenamefont {Webman}\ \emph {et~al.}(1977)\citenamefont {Webman},
  \citenamefont {Jortner},\ and\ \citenamefont
  {Cohen}}]{webman_thermoelectric_1977}%
  \BibitemOpen
  \bibfield  {author} {\bibinfo {author} {\bibfnamefont {I.}~\bibnamefont
  {Webman}}, \bibinfo {author} {\bibfnamefont {J.}~\bibnamefont {Jortner}}, \
  and\ \bibinfo {author} {\bibfnamefont {M.~H.}\ \bibnamefont {Cohen}},\
  }\bibfield  {title} {\emph {\bibinfo {title} {Thermoelectric power in
  inhomogeneous materials},\ }}\href {\doibase 10.1103/PhysRevB.16.2959}
  {\bibfield  {journal} {\bibinfo  {journal} {Physical Review B}\ }\textbf
  {\bibinfo {volume} {16}},\ \bibinfo {pages} {2959} (\bibinfo {year}
  {1977})}\BibitemShut {NoStop}%
\bibitem [{\citenamefont {Efimkin}\ and\ \citenamefont
  {Lozovik}(2013)}]{efimkin_drag_2013}%
  \BibitemOpen
  \bibfield  {author} {\bibinfo {author} {\bibfnamefont {D.~K.}\ \bibnamefont
  {Efimkin}}\ and\ \bibinfo {author} {\bibfnamefont {Y.~E.}\ \bibnamefont
  {Lozovik}},\ }\bibfield  {title} {\emph {\bibinfo {title} {Drag effect and
  {Cooper} electron-hole pair fluctuations in a topological insulator film},\
  }}\href {\doibase 10.1103/PhysRevB.88.235420} {\bibfield  {journal} {\bibinfo
   {journal} {Physical Review B}\ }\textbf {\bibinfo {volume} {88}},\ \bibinfo
  {pages} {235420} (\bibinfo {year} {2013})}\BibitemShut {NoStop}%
\bibitem [{\citenamefont {Gamucci}\ \emph {et~al.}(2014)\citenamefont
  {Gamucci}, \citenamefont {Spirito}, \citenamefont {Carrega}, \citenamefont
  {Karmakar}, \citenamefont {Lombardo}, \citenamefont {Bruna}, \citenamefont
  {Pfeiffer}, \citenamefont {West}, \citenamefont {Ferrari}, \citenamefont
  {Polini},\ and\ \citenamefont {Pellegrini}}]{gamucci_anomalous_2014}%
  \BibitemOpen
  \bibfield  {author} {\bibinfo {author} {\bibfnamefont {A.}~\bibnamefont
  {Gamucci}}, \bibinfo {author} {\bibfnamefont {D.}~\bibnamefont {Spirito}},
  \bibinfo {author} {\bibfnamefont {M.}~\bibnamefont {Carrega}}, \bibinfo
  {author} {\bibfnamefont {B.}~\bibnamefont {Karmakar}}, \bibinfo {author}
  {\bibfnamefont {A.}~\bibnamefont {Lombardo}}, \bibinfo {author}
  {\bibfnamefont {M.}~\bibnamefont {Bruna}}, \bibinfo {author} {\bibfnamefont
  {L.~N.}\ \bibnamefont {Pfeiffer}}, \bibinfo {author} {\bibfnamefont {K.~W.}\
  \bibnamefont {West}}, \bibinfo {author} {\bibfnamefont {A.~C.}\ \bibnamefont
  {Ferrari}}, \bibinfo {author} {\bibfnamefont {M.}~\bibnamefont {Polini}}, \
  and\ \bibinfo {author} {\bibfnamefont {V.}~\bibnamefont {Pellegrini}},\
  }\bibfield  {title} {\emph {\bibinfo {title} {Anomalous low-temperature
  {Coulomb} drag in graphene-{GaAs} heterostructures},\ }}\href {\doibase
  10.1038/ncomms6824} {\bibfield  {journal} {\bibinfo  {journal} {Nature
  Communications}\ }\textbf {\bibinfo {volume} {5}},\ \bibinfo {pages} {5824}
  (\bibinfo {year} {2014})}\BibitemShut {NoStop}%
\bibitem [{\citenamefont {Rodriguez-Vega}\ \emph {et~al.}(2014)\citenamefont
  {Rodriguez-Vega}, \citenamefont {Fischer}, \citenamefont {Das~Sarma},\ and\
  \citenamefont {Rossi}}]{rodriguez-vega_ground_2014}%
  \BibitemOpen
  \bibfield  {author} {\bibinfo {author} {\bibfnamefont {M.}~\bibnamefont
  {Rodriguez-Vega}}, \bibinfo {author} {\bibfnamefont {J.}~\bibnamefont
  {Fischer}}, \bibinfo {author} {\bibfnamefont {S.}~\bibnamefont {Das~Sarma}},
  \ and\ \bibinfo {author} {\bibfnamefont {E.}~\bibnamefont {Rossi}},\
  }\bibfield  {title} {\emph {\bibinfo {title} {Ground state of graphene
  heterostructures in the presence of random charged impurities},\ }}\href
  {\doibase 10.1103/PhysRevB.90.035406} {\bibfield  {journal} {\bibinfo
  {journal} {Physical Review B}\ }\textbf {\bibinfo {volume} {90}},\ \bibinfo
  {pages} {035406} (\bibinfo {year} {2014})}\BibitemShut {NoStop}%
\bibitem [{\citenamefont {Casimir}(1945)}]{casimir_onsagers_1945}%
  \BibitemOpen
  \bibfield  {author} {\bibinfo {author} {\bibfnamefont {H.~B.~G.}\
  \bibnamefont {Casimir}},\ }\bibfield  {title} {\emph {\bibinfo {title} {On
  {Onsager}'s {Principle} of {Microscopic} {Reversibility}},\ }}\href {\doibase
  10.1103/RevModPhys.17.343} {\bibfield  {journal} {\bibinfo  {journal}
  {Reviews of Modern Physics}\ }\textbf {\bibinfo {volume} {17}},\ \bibinfo
  {pages} {343} (\bibinfo {year} {1945})}\BibitemShut {NoStop}%
\end{thebibliography}%

\end{document}